\documentclass[fleqn,usenatbib,useAMS,twocolumn]{aastex62}

\usepackage{graphicx}
\usepackage{indentfirst}
\usepackage{latexsym}
\usepackage{multirow}
\usepackage{tabls}
\usepackage{epsfig}
\usepackage{multirow}
\usepackage{subfigure}
\usepackage{orcid}
\usepackage{comment}
\usepackage{hyperref}
\usepackage{color}
\usepackage{amssymb}
\usepackage{amsfonts}
\usepackage{amsmath}
\usepackage{bm}
\usepackage{mathrsfs}
\usepackage{dsfont}
\usepackage{acronym}

\acrodef{GW}[GW]{gravitational wave}
\acrodef{BNS}[BNS]{binary neutron star}
\acrodef{H0}[$H_0$]{the Hubble constant}
\acrodef{EM}[EM]{electromagnetic}
\acrodef{CBC}[CBC]{compact binary coalescence}

\newcommand{\de}{{\rm d}}

\def\eg{{\emph{e.g.~}}}

\shorttitle{The Hitchhiker's guide to the galaxy catalog approach for gravitational wave cosmology}
\shortauthors{Gair, Ghosh, Gray, Holz, Mastrogiovanni, Mukherjee, Palmese, Tamanini, et al.}

\begin{document}

%\title{Dark Standard Sirens 101: A guide to the galaxy catalog approach for gravitational wave cosmology}
\title{The Hitchhiker's guide to the galaxy catalog approach for dark siren gravitational wave-cosmology}

\author{Jonathan R. Gair
\orcidlink{0000-0002-1671-3668}}
\email{jgair@aei.mpg.de}
\affiliation{Max Planck Institute for Gravitationsphysik (Albert Einstein Institute), Am M\"{u}hlenberg 1, 14476 Potsdam, Germany}

\author{Archisman Ghosh \orcidlink{0000-0003-0423-3533}}
\email{archisman.ghosh@ugent.be}
\affiliation{Department of Physics and Astronomy, Ghent University, Proeftuinstraat 86, 9000 Ghent, Belgium}

\author{Rachel Gray \orcidlink{0000-0002-5556-9873}}
\email{r.gray@qmul.ac.uk}
\affiliation{Department of Physics and Astronomy, Queen Mary University of London, Mile End Road, London, E1 4NS, United Kingdom} 

\author{Daniel E. Holz
\orcidlink{0000-0002-0175-5064}}
%\email{}
\affiliation{Department of Physics, Department of Astronomy \& Astrophysics, The University of Chicago, 5640 South Ellis Avenue, Chicago, Illinois 60637, USA}

\author{Simone Mastrogiovanni \orcidlink{0000-0003-1606-4183}}
\email{mastrosi@roma1.infn.it}
\affiliation{INFN, Sezione di Roma, I-00185 Roma, Italy}

\author{Suvodip Mukherjee \orcidlink{0000-0002-3373-5236}
}
\email{suvodip@tifr.res.in}
\affiliation{Tata Institute of Fundamental Research, Homi  Bhabha Road, Mumbai, 400005, India}

\author{Antonella Palmese
\orcidlink{0000-0002-6011-0530}}
\email{palmese@cmu.edu}
\thanks{NASA Einstein Fellow}
\affiliation{McWilliams Center for Cosmology, Department of Physics, Carnegie Mellon University, Pittsburgh, PA 15213}
\affiliation{Department of Physics, University of California Berkeley, 366 LeConte Hall MC 7300, Berkeley, CA, 94720, USA}

\author{Nicola Tamanini}
\email{nicola.tamanini@l2it.in2p3.fr}
\affiliation{Laboratoire des 2 Infinis - Toulouse (L2IT-IN2P3), Universit\'e de Toulouse, CNRS, UPS, F-31062 Toulouse Cedex 9, France}

\author{Tessa Baker \orcidlink{0000-0001-5470-7616}}
%\email{t.baker@qmul.ac.uk}
\affiliation{Queen Mary University of London, Mile End Road, London E1 4NS, UK}

\author{Freija Beirnaert \orcidlink{0000-0002-4003-7233}}
%\email{freija.beirnaerti@ugent.be }
\affiliation{Department of Physics and Astronomy, Ghent University, Proeftuinstraat 86, 9000 Ghent, Belgium}

\author{Maciej Bilicki  \orcidlink{0000-0002-3910-5809}}
%\email{bilicki@cft.edu.pl}
\affiliation{Center for Theoretical Physics, Polish Academy of Sciences, al. Lotnik\'{o}w 32/46, 02-668 Warsaw, Poland}

\author{Hsin-Yu Chen \orcidlink{0000-0001-5403-3762}}
%\email{hsinyu@austin.utexas.edu}
\affiliation{Department of Physics, The University of Texas at Austin, Robert A Welch Hall, 105 E 24th st, Austin, TX 78712, USA}
\affiliation{Department of Physics and Kavli Institute for Astrophysics and Space Research, Massachusetts Institute of Technology, 77 Massachusetts Ave, Cambridge, MA 02139, USA}

\author{Gergely D\'{a}lya \orcidlink{0000-0003-3258-5763}}
%\email{gergely.dalya@ugent.be}
\affiliation{Department of Physics and Astronomy, Ghent University, Proeftuinstraat 86, 9000 Ghent, Belgium}

\author{Jose Maria Ezquiaga \orcidlink{0000-0002-7213-3211}}
%\email{jose.ezquiaga@nbi.ku.dk}
\affiliation{Niels Bohr International Academy, Niels Bohr Institute, Blegdamsvej 17, DK-2100 Copenhagen, Denmark}

\author{Will M. Farr \orcidlink{0000-0003-1540-8562}}
%\email{will.farr@stonybrook.edu}
\affiliation{Department of Physics and Astronomy, Stony Brook University, Stony Brook NY 11794, USA}
\affiliation{Center for Computational Astrophysics, Flatiron Institute, New York NY 10010, USA}

\author{Maya Fishbach \orcidlink{0000-0002-1980-5293}}
%\email{fishbach@cita.utoronto.ca}
\affiliation{Canadian Institute for Theoretical Astrophysics, David A. Dunlap Department of
Astronomy and Astrophysics, and Department of Physics, 60 St George St, University of Toronto, Toronto, ON M5S 3H8, Canada}

\author{Juan Garcia-Bellido \orcidlink{0000-0002-9370-8360}}
%\email{juan.garciabellido@uam.es}
\affiliation{Instituto de Física Teórica UAM/CSIC, Universidad Autónoma Madrid, 28049 Madrid, Spain}

\author{Tathagata Ghosh \orcidlink{0000-0001-9848-9905}}
%\email{tathag@iucaa.in}
\affiliation{Inter-University Centre for Astronomy and Astrophysics, Post Bag 4, Ganeshkhind, Pune 411 007, India}

\author{Hsiang-Yu Huang \orcidlink{0000-0002-1665-2383}}
%\email{hyhuanggw170817@gate.sinica.edu.tw}
\affiliation{Department of Physics, National Central University, No. 300, Zhongda Rd., Zhongli District, Taoyuan City 320317, Taiwan}

\author{Christos Karathanasis \orcidlink{0000-0002-0642-5507}}
%\email{ckarathanasis@ifae.es}
\affiliation{Institut de Física d’Altes Energies (IFAE), Universitat Autònoma de Barcelona, 
08193, Barcelona, Spain}

\author{Konstantin Leyde \orcidlink{0000-0001-7661-2810}}
%\email{kleyde@apc.in2p3.fr}
\affiliation{Universit\'e Paris Cit\'e, CNRS, Astroparticule et Cosmologie, F-75006 Paris, France}

\author{Ignacio Maga\~na Hernandez \orcidlink{0000-0003-2362-0459}}
%\email{maganah2@uwm.edu}
\affiliation{University of Wisconsin-Milwaukee, Milwaukee, WI 53201, USA}

\author{Johannes Noller \orcidlink{0000-0003-2210-775X}}
%\email{johannes.noller@port.ac.uk}
\affiliation{Institute of Cosmology \& Gravitation, University of Portsmouth, Portsmouth, PO1 3FX, U.K.}
\affiliation{DAMTP, University of Cambridge, Wilberforce Road, Cambridge CB3 0WA, U.K.}

\author{Gregoire Pierra \orcidlink{0000-0003-3970-7970}}
%\email{g.pierra@ip2i.in2p3.fr}
\affiliation{Université Claude Bernard Lyon 1, CNRS, Institut de physique des deux infinis, 4 rue Enrico Fermi, 69100 Villeurbanne, France}

\author{Peter Raffai \orcidlink{0000-0001-7576-0141}}
%\email{peter.raffai@ttk.elte.hu}
\affiliation{Institute of Physics, E\"otv\"os Lor\'and University, 1117 Budapest, Hungary}

\author{Antonio Enea Romano \orcidlink{0000-0002-0314-8698}}
%\email{antonioenea.romano@ligo.org}
\affiliation{Instituto de F\'isica, Universidad de Antioquia, A.A.1226, Medell\'in, Colombia}
\affiliation{ICRANet, Piazza della Repubblica 10, I--65122 Pescara, Italy}

\author{Monica Seglar-Arroyo \orcidlink{0000-0001-8654-409X}}
%\email{mseglar@ifae.es}
\affiliation{Institut de Física d’Altes Energies (IFAE), Universitat Autònoma de Barcelona, 
08193, Barcelona, Spain}

\author{Dani\`ele A.~Steer \orcidlink{0000-0002-8781-1273}}
%\email{steer@apc.in2p3.fr}
\affiliation{Universit\'e Paris Cit\'e, CNRS, Astroparticule et Cosmologie, F-75013 Paris, France and Laboratoire de Physique de l'\'Ecole Normale Su\'eperieure, ENS, CNRS, Universit\'e PSL, Sorbonne Universit\'e, Universit\'e Paris Cit\'e, F-75005 Paris, France}

\author{Cezary Turski \orcidlink{0000-0001-7122-6240}}
%\email{cezary.turski@ugent.be }
\affiliation{Department of Physics and Astronomy, Ghent University, Proeftuinstraat 86, 9000 Ghent, Belgium}

\author{Maria Paola Vaccaro \orcidlink{0000-0003-3776-9246}}
%\email{mvaccaro@ifae.es}
\affiliation{Institut de Física d’Altes Energies (IFAE), Universitat Autònoma de Barcelona, 
08193, Barcelona, Spain}

\author{Sergio Andr\'es Vallejo-Pe\~na  \orcidlink{0000-0002-6827-9509}}
%\email{sergio.vallejo@udea.edu.co}
\affiliation{Instituto de F\'isica, Universidad de Antioquia, A.A.1226, Medell\'in, Colombia}

\keywords{catalogs --- cosmology: observations --- gravitational waves --- surveys}

\date\today

\begin{abstract}
We outline the ``dark siren'' galaxy catalog method for cosmological inference using gravitational wave (GW) standard sirens, clarifying some common misconceptions in the implementation of this method. When a confident  transient electromagnetic counterpart to a GW event is unavailable, the identification of a unique host galaxy is in general challenging. Instead,  as originally proposed by~\citet{schutz}, one can
consult a galaxy catalog and implement a dark siren statistical approach incorporating all potential host galaxies within the localization volume. \citet{2021arXiv211200241T} recently claimed that this approach results in a biased estimate of the Hubble constant, $H_0$, when implemented on mock data, even if optimistic assumptions are made. We demonstrate explicitly that, as previously shown by multiple independent groups, the dark siren statistical method leads to an unbiased posterior when the method is applied to the data correctly. 
We highlight common sources of error possible to make in the generation of mock data and implementation of the statistical framework, including the mismodeling of selection effects and inconsistent implementations of the Bayesian framework, which can lead to a spurious bias.
\end{abstract}

\section{Introduction}

The use of \acp{GW} from compact binary mergers as standard sirens \citep{schutz,2005ApJ...629...15H,Dalal:2006qt} for cosmology is an idea which has finally come to fruition in recent years. These signals directly provide a measurement of the luminosity distance measurement to the source, which is therefore independent of the cosmic distance ladder. With the addition of redshift information, measurements can therefore be made of those cosmological parameters which impact the expansion history of the Universe, such as \ac{H0}. This approach is independent of all other local measurements to date. 

The detection of the \ac{BNS} GW170817 \citep{ligobns} and its \ac{EM} counterpart (\citealt{MMApaper}) by the LIGO \citep{LIGOScientific:2014pky} and Virgo \citep{VIRGO:2014yos} GW detectors --- which allowed the host galaxy, and hence the redshift of the merger, to be identified --- led to the first \ac{GW} measurement of \ac{H0} \citep{2017Natur.551...85A}. In the absence of an \ac{EM} counterpart, redshift information from other sources can be used, such as i) galaxy catalogs (using the statistical method \citep{PhysRevD.86.043011,chen17,fishbach,2020PhRvD.101l2001G,Gray2022,PhysRevD.105.023523} or the cross-correlation method exploring the spatial clustering between GW sources and galaxies \citep{PhysRevD.93.083511,Mukherjee:2019wcg, Mukherjee:2020hyn,Bera:2020jhx, Diaz:2021pem}) and ii) ``spectral siren'' inference of the redshift from the astrophysical distributions of the \ac{GW} sources themselves
\citep{1993ApJ...411L...5C,Taylor_2012,Farr_2019,2020arXiv200602211M,mastrogiovanni_2021,Mukherjee:2021rtw,2022JCAP...09..012L,Ezquiaga_2022, Karathanasis:2022rtr}.

This paper will focus on the \emph{dark siren + galaxy catalog} method (also informally known as the \emph{statistical} or \emph{dark siren} method),\footnote{In general, any \ac{GW} observed without an \ac{EM} counterpart is a dark siren, and there is more than one method which can be used to produce cosmological measurements with dark sirens, so it is worth making the distinction. Similarly, all \ac{GW} measurements of \ac{H0} are statistical in that they require the information from many \ac{GW} events to be combined in order to make a constraint.} in which galaxy catalogs are used to provide the redshift information for all potential host galaxies within a \ac{GW}'s localisation volume, which are statistically averaged over. While less informative than the counterpart method on an event-by-event basis, the constraint strengthens as more events are included in the analysis.  Given the current detection rates of bright and dark sirens (the latter have a factor $>10$ more detections), this method is expected to make a significant contribution to the \ac{GW} constraint of \ac{H0}~\citep{chen17}. Indeed, this approach has already been implemented
%many works have already applied this method
using dozens of available GW detections \citep{fishbach,darksiren1,LVC_O2_StS,palmese20_sts,gwtc3_H0,finke2021cosmology,2021arXiv211106445P}.   

This paper aims to act as a point of introduction for those new to the field of \ac{GW} cosmology, and the dark siren + galaxy catalog method in particular, using mock data to build up a basic analysis following a Bayesian approach. This is motivated by the recent claim that GW dark sirens generally provide biased estimates of $H_0$ even under simplified assumptions (\citealt{2021arXiv211200241T}; TH21 hereinafter). We show here that this is not the case, and how it is incorrect modeling assumptions that lead to biased measurements. 
Given this, it is important to stress that a lot of the inconsistencies and biases explored in this work are not relevant for the case of realistic data, and instead primarily arise due to simplified assumptions and toy models for the GW sources and galaxy catalogs.

It is well understood that biases cannot arise in statistical analysis when the model and data-generating process are consistent. In simulation studies, it is possible to control the data generation process and so if biases appear these must be due to errors or inconsistencies when generating or analyzing the data. This should be used as a diagnostic tool to track down those errors and inconsistencies. This does not mean that analyses of real data are free of biases, since the true data-generating process for any observed process is unknowable. However, the true parameters are also not then known and so biases are hard to diagnose. It is certainly valuable, within a simulation study, to vary the assumptions about the data-generating process, while not changing the analysis, to understand what biases could appear in the analysis of real data. Indeed, this is the correct procedure for identifying potential sources of systematic error. However, this is only useful if a consistent analysis has first been shown to give unbiased results, and if the types of modifications to the data-generating process are controlled and physically well-motivated. Any sources of bias identified in this way can then be mitigated by increasing the complexity of the analysis model.

This paper is organised as follows. Section \ref{sec:method} introduces the basic Bayesian framework, then describes the mathematics of how the mock \ac{GW} and \ac{EM} galaxy catalogue data should be constructed. Section \ref{sec:results} gives details of the simulated data, then shows results for the scenario in which 200 \ac{GW} detections are used to constrain \ac{H0}, and how the constraint changes in the limit of a large number of detections. Section \ref{sec:discussion} discusses some of the common mistakes made when applying the dark siren + galaxy catalog method, particularly when generating simplified mock data, which can lead to biased outcomes. It also highlights some of the real-world complications which need to be addressed in a true dark siren + galaxy catalog analysis, which are not in scope for this paper.
We note that this work is {\bf not} intended to present a forecast of how well we will be able to constrain $H_0$, as we make some simplifying assumptions that are different from reality, but it is intended as a pedagogical work on the method.  %

\section{Statistical framework \label{sec:method}}

Let us assume that we have observed a set of $N_{\rm obs}$ GW events $\{x\}$
%$\{\rm GW\}$ 
from which we are able to measure the luminosity distance of each source.  In this basic example, $H_0$ can be determined by the fact that we measure the source luminosity distance from the GWs and can identify potential host galaxies from the catalog.
According to Bayes' theorem, the posterior on $H_0$, given a set of detected GW events $\{x\}$, can be written as
\begin{equation}
    p(H_0|\{x\}) \propto \mathcal{L}(\{x\}|H_0) p(H_0),
\end{equation}
where $p(H_0)$ is the prior on $H_0$. The likelihood $\mathcal{L}(\{x\}|H_0)$ can be described by an inhomogeneous Poisson process \citep{mandel,2022hgwa.bookE..45V} 
\begin{eqnarray}
    \mathcal{L}(\{x\}&|&H_0) = \frac{e^{-N_{\rm exp}(H_0)}~ [N_{\rm exp}(H_0)]^{N_{\rm obs}}}{N_{\rm obs}!}  \nonumber \\ &\times&  \prod_{i}^{N_{\rm obs}} \frac{\int \de z~ \mathcal{L}_{\rm GW}(x_i|d_L(z,H_0))~p_{\rm CBC}(z)}{\int \de z~ P^{\rm GW}_{\rm det}(z,H_0) ~p_{\rm CBC}(z)}.
\end{eqnarray}
Here $\mathcal{L}_{\rm GW}(x_i|d_L(z,H_0))$ is the GW likelihood %(how well you measure the luminosity distance)
(the probability of measuring the data $x_i$ given some luminosity distance $d_L$), $p_{\rm CBC}(z)$ is the probability that the source, a \acp{CBC}, is at redshift $z$ for an observer on Earth, while  $p_{\rm det}^{\rm GW}(z, H_0)$  is the GW  detection probability used to account for selection biases. The term $N_{\rm exp}(H_0)$ is the number of expected GW detections for a given value of $H_0$. This term can be marginalized analytically by assuming a $1/N_{\rm exp}$ prior on the expected number of events \citep{2018ApJ...863L..41F}, or equivalently the compact object binary merger rate. With this assumption, the pre-factor loses its dependence on \ac{H0} and the likelihood for a single event reduces to 
\begin{equation}
    \mathcal{L}(x_i|H_0) =  \frac{\int \de z~ \mathcal{L}_{\rm GW}(x_i|d_L(z,H_0))~p_{\rm CBC}(z)}{\int \de z~ P^{\rm GW}_{\rm det}(z,H_0)~ p_{\rm CBC}(z)},
    \label{eq:hierarchica}
\end{equation}
which is the one we use here. The quantity $p_{\rm CBC}(z)$ is being used here to represent the distribution of redshifts from which the GW source is drawn. If this is known, the likelihood, which is the probability distribution of observed data sets, can be found by marginalising over the distribution of possible redshifts, which is what is done in Eq.~(\ref{eq:hierarchica}). In practice we do not know $p_{\rm CBC}(z)$, but we construct it from observations of galaxies, as described in section~\ref{sec:galcat} below. The interpretation is then that we are using the \emph{posterior} from EM observations as the \emph{prior} for the GW data. This is perfectly legitimate, but to be rigorously correct, what the EM data provides is a distribution of possible GW source distributions, since the EM data is not perfect. In practice, this means that we should treat the unknown true redshifts of the galaxies as population level parameters that we infer jointly from the EM and GW data, and marginalise over these after combining the individual GW measurements. In simpler terms the integral over $z$ in Eq.~(\ref{eq:hierarchica}) should be done at the end, rather than separately for each event. However, as we discuss more in detail in Sec.~\ref{sec:more}, it can be shown that in the limit that the redshifts of the galaxies are perfectly known or the number density of CBCs is much lower than the number density of galaxies the hierarchical likelihood for $H_0$ can be reduced to Eq.~(\ref{eq:hierarchica}). In particular, the second hypothesis is perfectly reasonable considering that currently the rate of \ac{CBC} mergers is estimated to be $\sim 10^{-6}$--$10^{-5} \, {\rm yr}^{-1}$ per galaxy \citep{LIGOScientific:2020kqk}. For this reason,  Eq.~(\ref{eq:hierarchica}) is used in most current analyses and this is perfectly legitimate. Where this distinction can matter, however, is in a mock data scenario where the number of CBCs is artificially inflated (or the density of galaxies artificially reduced). For most of this paper, we will limit our discussion to the case in which the number density of CBCs is much lower than the number density of galaxies. In sections~\ref{sec:more} and~\ref{sec:onegal} only we will demonstrate when this assumption has an impact and how this can be mitigated.

%To be rigorously correct, for the hierarchical inference, we should not use a single redshift $z$, but many different and independent redshifts $z_i$ that represent the \textit{true} redshifts of all the galaxies in the universe. In fact, since galaxies are in common between all the GW events, in principle, the true redshift of all the galaxies should be treated as population-level parameters in a similar way to $H_0$. In the case that the \textit{true} redshifts of all the galaxies are not known, the marginalization over redshift in Eq. \ref{eq:hierarchica} should be done jointly over all the \textit{true} redshifts of the galaxies.  

As a final remark, it is important to note that the form of Eq.~(\ref{eq:hierarchica}) is based on the usual assumption that ``detection'' is a property of the observed data only, not the true parameters of the source. It can be convenient to simulate data by selecting events based on the true source parameters, but this is a modification to the \emph{prior} rather than to the \emph{likelihood} ad would be inconsistent with the detection model assumed in Eq.~(\ref{eq:hierarchica}). A consistent analysis in this case would just retain the numerator of Eq.~(\ref{eq:hierarchica}), but with the prior $p_{\rm CBC}(z)$ renormalised by dividing by its integral over the range of events that are detectable (which will be dependent on the cosmological parameters) and the integral in the numerator truncated to the same range. While this correction should eliminate biases, we would not advocate this approach as it does not reflect the reality of an analysis of real data and might therefore give misleading results.

\subsection{Galaxy Catalog modeling}
\label{sec:galcat}
In this section we describe how to build the probability $p_{\rm CBC}(z)$ of a \ac{CBC} occurring at redshift $z$, under the assumption that \acp{CBC} occur in galaxies, and so will trace the distribution of galaxies in the universe in some fashion. Note that in this section (and for the rest of the paper) we will discuss the distribution of galaxies purely as a function of redshift, and thus neglect their spatial distribution in right ascension and declination. This closely follows the assumptions and approximations made in TH21. Neglecting this aspect does not automatically introduce problems, but it is an unrealistic set-up which increases the risk mentioned above about relative number-densities of \acp{CBC} and galaxies.

The probability that a merger will occur at a redshift $z$ is the product of the probability that \textit{there is a galaxy at $z$}, $p_{\rm cat}(z)$, and the probability of a galaxy at redshift $z$ \textit{hosting a \ac{GW} merger}, $p_{\rm rate}(z)$,
\begin{equation}
  p_{\rm CBC}(z) = \frac{p_{\rm rate}(z) p_{\rm cat}(z)}{\int_0^\infty p_{\rm rate}(z) p_{\rm cat}(z) \de z}.  
  \label{eq:pcbc}
\end{equation}
The rate -- or signal emission -- probability is given by
\begin{equation}
    p_{\rm rate}(z) = \frac{\frac{R(z)}{1+z}}{\int_0^{\infty} \frac{R(z)}{1+z} \de z},
    \label{eq:rate}
\end{equation}
where $R(z)$ is the merger rate of GWs in their source frame usually expressed in ${\rm Gpc^{-3}yr^{-1}}$. The merger rate is usually parametrised as $R(z) \propto (1+z)^\gamma$ in the redshift region $0<z<2$ \citep{2017ApJ...851L..25F} to account for a possible evolution of the merger rate.
If GW mergers are uniform in comoving volume and source-frame time, this term is constant. The $1/1+z$ factor accounts for the effect of time dilation due to the expansion of the Universe between the source frame and the detector frame. In absence of any galaxy catalog observation, $p_{\rm cat}(z)$ can be constructed with a uniform in comoving volume distribution and 
\begin{equation}
    p_{\rm CBC}(z) = \frac{\frac{R(z)}{1+z} \frac{dV_c}{dz}}{\int \frac{R(z)}{1+z} \frac{dV_c}{dz} \de z},
    \label{eq:pppp2}
\end{equation}
where 
\begin{equation}
    \frac{dV_c}{dz} = 4 \pi \left[\frac{c}{H_0}\right]^3 \left[\int_0^z \frac{\de z'}{E(z')}\right]^2.
\end{equation}
Note that the above prior does not depend on $H_0$, but it depends on other cosmological parameters through the expansion history ($H(z)/H_0 \equiv E(z)= [\Omega_m(1+z)^3+ \Omega_\Lambda]^{1/2}$ for a Flat Lambda Cold Dark Matter cosmological model) and the parametrization of the rate term. This is the prior used for analyses making use of mass information \citep{1993ApJ...411L...5C,Taylor_2012,Farr_2019,mastrogiovanni_2021,Mukherjee:2021rtw, 2022JCAP...09..012L,Ezquiaga_2022, Karathanasis:2022rtr}, where cosmology and $p_{\rm CBC}(z)$ are fit jointly. Note also that if the CBC rate presents some features in redshift, such as peaks, it might help to measure cosmology even in absence of counterparts or galaxy information \citep{2021PhRvD.104d3507Y}.

. Physically, $p_{\rm CBC}(z)$ is something like the redshift distribution of galaxies that have sourced a compact merger, the signal of which has passed through the Earth.
It is interesting to note that Eq.~\ref{eq:pppp2} can be also defined from a more ``physical'' argument starting from the rate of CBC mergers seen from an observer of Earth. This quantity can be expressed as
\begin{eqnarray}
    \frac{\de N_{\rm CBC}}{\de z \de t_d}&=& \frac{\de N_{\rm CBC}}{\de V_c \de t_s} \frac{\de V_c}{\de z} \frac{\de t_s}{\de t_d} \nonumber \\ 
    &=& R(z) \frac{\de V_c}{\de z} \frac{1}{1+z},
\end{eqnarray}
where $t_d$ and $t_s$ are the times in the detector and source frame and the number of CBC per comoving volume per source frame time is by definition the CBC merger rate.

For this basic mock data, when calculating Eq.~\ref{eq:pcbc}, we will neglect the rate term in Eq.~\ref{eq:rate} in order to align more closely with the approach taken in TH21. As long as this rate assumption is treated self-consistently when generating the mock data and analyzing it, this will not introduce any bias to the results. Let us note that this is a simplified description where CBC rates only depend on the Universe epoch (redshift), in a more general case, we might even model that more luminous galaxies are more likely to host CBCs. Therefore, in the remaining of the paper, we will approximate
\begin{equation}
    p_{\rm CBC}(z) \approx p_{\rm cat}(z).
\end{equation}

Moreover, here we want to exploit the fact that we are provided with a galaxy survey.  We want to build $p_{\rm cat}(z)$ given the observation of $N_{\rm gal}$  galaxies with \textit{measured redshifts} $\{\hat{z}_g\}$. In other words, now we are computing $p_{\rm cat}(z|\{\hat{z}_g\})$. In this computation, we will assume that the galaxy catalog is complete.\footnote{``Complete'' is defined here to mean that the catalog contains all galaxies that can emit \acp{CBC} in the universe.} For details on how galaxy catalog incompleteness can be incorporated into such an analysis, see \eg \cite{chen17,fishbach,2020PhRvD.101l2001G,Gray2022}. While constructing $p_{\rm cat}(z|\{\hat{z}_g\})$, we must consider that $\{\hat{z}_g\}$ are not the \textit{true} redshifts $\{z_g\}$. We can take into account this uncertainty using the laws of probabilities, namely
\begin{equation}
  p_{\rm cat}(z|\{\hat{z}_g\})=\int \de \{z_g\}p_{\rm gal}(z|\{z_g\})p_{\rm red}(\{z_g\}|\{\hat{z}_g\}),
  \label{eq:pgal}
\end{equation}
where $p_{\rm gal}(z|\{z_g\})$ is the probability to have a galaxy at redshift $z$ when we  have a set of true redshifts for the galaxies. This is simply given by
\begin{equation}
    p_{\rm gal}(z|\{z_g\}) = \frac{1}{N_{\rm gal}}\sum_i^{N_{\rm gal}} \delta(z-z_g^i),
    \label{eq:delta}
\end{equation}
where $\delta$ is  Dirac delta distribution. The probability $p_{\rm red}(\{z_g\}|\{\hat{z}_g\})$ encodes the fact that we are not perfectly able to measure the true redshift of the galaxies, but we have a \textit{measured value}. We label this probability with ``red'' to indicate that it refers to the measurement uncertainties due to redshift. Since each galaxy measure is independent of the others,
\begin{equation}
  p_{\rm red}(\{z_g\}|\{\hat{z}_g\})=\prod_j p_{\rm red}(z_g^j|\hat{z}_g^j).  
  \label{eq:redunc}
\end{equation}
By plugging Eqs.~\ref{eq:delta}--\ref{eq:redunc} in Eq.~\ref{eq:pgal} we obtain that
\begin{equation}
  p_{\rm CBC}(z) \approx p_{\rm cat}(z|\{\hat{z}_g\}) = \frac{1}{N_{\rm gal}}  \sum_{i}^{N_{\rm gal}} p_{\rm red}(z|\hat{z}_g^i).
  \label{eq:pofzc}
\end{equation}
At this point we can immediately note that if the galaxy redshifts are perfectly measured, Eq.~\ref{eq:pofzc} reduces to a sum of Dirac delta functions
\begin{equation}
  p_{\rm CBC}(z) \approx p_{\rm cat}(z|\{\hat{z}_g\}) = \frac{1}{N_{\rm gal}}  \sum_{i}^{N_{\rm gal}} \delta (z-\hat{z}_g^i)
  \label{eq:pofzc2},
\end{equation}
and the likelihood in Eq.~\ref{eq:hierarchica} can be expressed analytically as
\begin{equation}\label{Eq:liknouncert}
    \mathcal{L}(x_i|H_0) =  \frac{\sum_i^{N_{\rm gal}}  \mathcal{L}_{\rm GW}(x_i|d_L(\hat{z}^i_g,H_0))}{\sum_i^{N_{\rm gal}} P^{\rm GW}_{\rm det}(\hat{z}^i_g,H_0)}.
\end{equation}
In Sec.~\ref{sec:more} we show that the hierarchical likelihood still reduces to this equation even if we relax the assumption that the CBC density is lower than the galaxy density.

However, we are usually in a situation where the redshift of the galaxies is measured with large uncertainties. The term $p_{\rm CBC}(z)$ is our \emph{prior} on the \ac{CBC} redshift distribution. In order for it to be used as such, the sum in Eq. \ref{eq:pofzc} must be over the redshift posteriors of individual galaxies. In this case, we will construct the likelihood of the individual galaxies in the same manner as was used to construct the \ac{GW} likelihood, but an additional prior must be applied to convert this to a posterior.
That is:
\begin{equation}
    p_{\rm red}(z|\hat{z}_g^i) = \frac{\mathcal{L}_{\rm red}(\hat{z}^i_g|z) p_{\rm bg}(z)}{\int \mathcal{L}_{\rm red}(\hat{z}^i_g|z) p_{\rm bg}(z) \de z },
    \label{eq:redpos}
\end{equation}
where $\mathcal{L}_{\rm red}(\hat{z}^i_g|z)$ is the likelihood model used to generate \textit{observed} redshifts from the true redshift of each galaxy, and $p_{\rm bg}(z)$ is a chosen prior on redshift that reflects our best belief on the background distribution of galaxies.
We adopt a Gaussian redshift likelihood model like in TH21:
\begin{equation}
    \mathcal{L}(\hat{z}^i_g|z)=\frac{1}{\sqrt{2\pi}\sigma_z} e^{-\frac{(\hat{z}^i_g-z)^2}{2\sigma^2_z}},
    \label{eq:redlikeli}
\end{equation}
with $\sigma_z=0.013(1+z)^3 \leq 0.015$. Note that the redshift uncertainty explicitly depends on the \emph{true} redshift, and this has to be taken into account in the modeling. 

Regarding priors, in absence of any special knowledge about the galaxy distribution, the simplest choice we can make is to set $p_{\rm bg}(z)$ uniform in comoving volume. This can appear as ``double counting'' but it is the more conservative choice in absence of any data-driven information since we  would expect galaxies to be nearly uniformly distributed in comoving volume. In the limit that $\sigma_z \rightarrow \infty $, i.e. we detect galaxies but we do not measure their redshifts, $p_{\rm CBC}(z)$ will then represent galaxies uniform in comoving volume. While in the limit that $\sigma_z \rightarrow 0$, the prior choice $p_{\rm bg}(z)$ will not matter and $p_{\rm CBC}(z)$ will be given by the sum of $\delta$-like peaks located at the measured redshift of galaxies.

\subsection{GW data modeling}

The problems of detection and source parameters estimation of GW signals are complex and an active field of current research. We refer the reader to \citet{2020CQGra..37e5002A} for a more in-depth discussion. Here we just discuss the fundamental aspects of detection and parameter estimation for GW signals. 

GW signals are detected using low-latency algorithms, able to calculate signal-to-noise ratios and false alarm rate using either a template bank \citep{2017PhRvD..95d2001M,2021SoftX..1400680C,2021CQGra..38i5004A,DalCanton:2014hxh,2021CQGra..38i5004A,2017ApJ...849..118N} or a superposition of wavelets \citep{Klimenko:2004qh}. Typically, GW signals with false alarm rate lower than 1--10 per year are selected for population analyses \citep{LIGOScientific:2020kqk,LIGOScientific:2021psn}. The choice of a detection threshold is important to correct for selection biases.
Once a signal is classified as detected, the source physical parameters are estimated using a Bayesian sampling of the GW likelihood:
\begin{equation}
    \mathcal{L}(x_i|\vec{\theta}) \propto e^{-\frac{1}{2}(x_i-h(\vec{\theta})|x_i-h(\vec{\theta}))},
\end{equation}
where $\vec{\theta}$ are the set of GW source parameters, $h$ is the GW signal, and $x_i$ indicates the GW strain data. The scalar product is given in Fourier frequency space by
\begin{equation}
    (a|b)=4 \mathrm{Re} \left[\int_0^{\infty} \frac{\tilde{a}(f) \tilde{b}^*(f)}{S(f)} \right],
\end{equation}
where $S(f)$ is the one-sided power spectral density of the noise. 

For this analysis, we do not use the full GW likelihood, and instead use a toy model similar to the one used in TH21. The GW likelihood in Eq.~\ref{eq:hierarchica} is a central quantity of the mock study. The GW likelihood provides a model to generate error budgets on the estimation of the luminosity distance and it is also used to define the detection probability 
\begin{equation}
 P^{\rm GW}_{\rm det}(z,H_0)= \int_{\rm detectable}  \mathcal{L}_{\rm GW}(x|d_L(z,H_0)) \de x,
 \label{eq:detprob}
\end{equation}
where the integral is done over all the possible detectable GW signals. In principle, the GW likelihood and detection probability should take into account all the GW parameters but in this simplified mock study, we will just model it as a function of $d_L$. We simplify the GW likelihood by defining an ``observed'' luminosity distance $\hat{d}_L^i$, which replaces the ``observed data'' $x_i$. We use a likelihood model 
\begin{equation}
    \mathcal{L}_{\rm GW}(\hat{d}^i_{L}|d_L(z,H_0))=\frac{1}{\sqrt{2\pi}\sigma_{d_L}} e^{-\frac{(\hat{d}^i_{L}-d_L(z,H_0))^2}{2\sigma^2_{d_L}}},
    \label{eq:GWlikeli}
\end{equation}
where $\sigma_{d_L}=A d_L(z,H_0)$, with $A$ a constant fractional error.  Note that Eq.~\ref{eq:GWlikeli} is a probability density function of the ``measured'' luminosity distance $\hat{d}^i_L$. In other words, given a value of the true luminosity distance $d_L$, the probability of obtaining a certain measured value $\hat{d}^i_L$ is distributed according to a normal distribution. However,  we note that the reconstruction of the ``true'' luminosity distance $d_L$ given an observed value $\hat{d}^i_L$ is \textbf{not} Gaussian.\footnote{This is due to the fact that there is a $d_L$ dependence in the denominator of Eq.~(\ref{eq:GWlikeli}) and \begin{equation*}
    p(d_L^i|\hat{d}^i_L)=\frac{\mathcal{L}(\hat{d}^i_L|d_L^i)\Pi(d_L^i)}{\int \mathcal{L}(\hat{d}^i_L|d_L^i)\Pi(d_L^i) \de d_L^i},
\end{equation*}
where $\Pi$ is a prior term.}
We also note that, while a simple likelihood model for $d_L$  is sometimes assumed in GW analyses (e.g. \citealt{Palmese_2021}) using the \texttt{bayestar} \citep{bayestar} sky maps, in general, this likelihood will take more complicated forms.

Regarding the detection process, we assume that GWs are detectable if their \emph{measured} luminosity distance is smaller than a threshold of $\hat{d}_L< \hat{d}^{\rm thr}_L= 1550$ Mpc. The GW detection probability in Eq.~\ref{eq:detprob} can be written using the likelihood model in Eq.~\ref{eq:GWlikeli} as
\begin{eqnarray}
    P^{\rm GW}_{\rm det}(z,H_0)&=& \int_{-\infty}^{\infty} \Theta(\hat{d}_{L};\hat{d}^{\rm thr}_L) \mathcal{L}_{\rm GW}(\hat{d}_{L}|d_L(z,H_0)) \de \hat{d}_L \nonumber \\ &=&
    \int_{-\infty}^{\hat{d}^{\rm thr}_L} \mathcal{L}_{\rm GW}(\hat{d}_{L}|d_L(z,H_0)) \de \hat{d}_L \nonumber \\ &=& \frac{1}{2}\left[1+{\rm erf}[\frac{d_L(z,H_0)-\hat{d}^{\rm thr}_L}{\sqrt{2}A d_L(z,H_0)}]\right].
    \label{eq:gwdet}
\end{eqnarray}
In the above Equations, $\Theta$ is a Heaviside step function of $\hat{d}_L$, dropping to 0 for $\hat{d}_L>\hat{d}^{\rm thr}_L$, ``erf'' is the unilateral error function of a standardized normal distribution.
In TH21 it is argued that $P^{\rm GW}_{\rm det}(z,H_0)$ is a Heaviside step function dropping to 0 at $\hat{d}^{\rm thr}_L$. This is correct only in the limit that $\sigma_{d_L} \rightarrow 0$ (for small error budgets).
\begin{figure}
    \centering
    \includegraphics{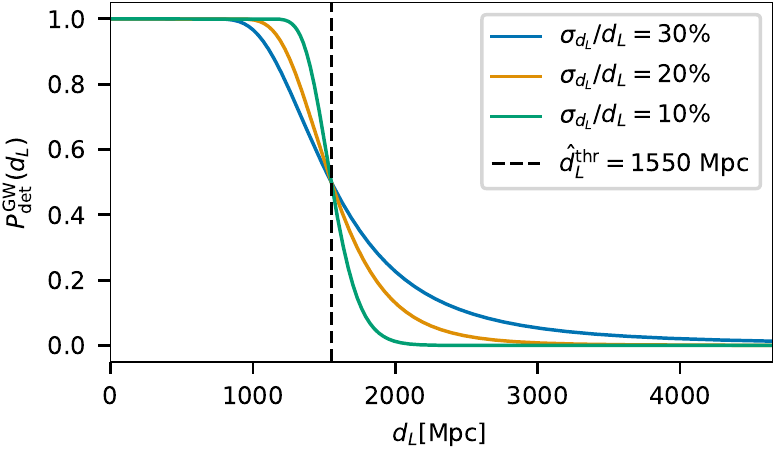}
    \caption{GW detection probability computed with the toy likelihood model in Eq.~\ref{eq:gwdet} as a function of the \textit{true} luminosity distance. The colors indicate various fractional errors on the luminosity distance. The vertical dashed line indicates the detection threshold for the \textit{measured} luminosity distance $\hat{d}_L$. We note that the detection probability only reduces to a Heaviside step function in the limit of $\sigma_{d_L}\rightarrow0$. If $\sigma_{d_L}\neq 0$, a step function on \emph{true} luminosity distance cannot be used to account for selection effects, else the $H_0$ posterior may be biased in a way that depends on $\sigma_{d_L}$.}
    \label{fig:GW_detection_prob}
\end{figure}
In Fig.~\ref{fig:GW_detection_prob} we plot the detection probability as a function of the true luminosity distance of the GW. When the GW likelihood has a non-negligible error, the GW detection probability does not drop sharply to zero. This is expected, as the scattering process of $d_L$ makes us able to detect sources whose true luminosity distance is above the detection threshold. The detection probability reduces to a Heaviside step function only when the distance error goes to 0 (i.e.~when we are perfectly able to measure luminosity distance).

\subsection{Full likelihood derivation}
\label{sec:more}
As described above, when writing down Eq.~(\ref{eq:hierarchica}) we have effectively assumed that the GW likelihood depends on a distribution $p_{\rm CBC}(z)$, which is known. We then proceeded to construct $p_{\rm CBC}(z)$ from EM observations, $\{ \hat{z}_g\}$, of galaxies in the catalogue via Eq.~(\ref{eq:pofzc}). Using the EM data as a prior for the GW data is perfectly legitimate, but the usual form of the GW likelihood, Eq.~(\ref{eq:hierarchica}) is then just an approximation. The GW likelihood actually depends on the true, and unknown, redshifts of the galaxies, $\{z_g\}$. In the absence of selection effects, this distinction does not matter for a single event, which we will now demonstrate. The GW likelihood for a single observation, $x_i$, is
\begin{equation}
{\cal L}_{\rm GW}(x_i|\{z_g\},H_0) = \frac{1}{N_{\rm gal}} \sum_{j=1}^{N_{\rm gal}} {\cal L}_{\rm GW}(x_i | d_L(z^j_g, H_0)),
\end{equation}
the likelihood for the EM observations is
\begin{equation}
{\cal L}_{\rm EM}(\{\hat{z}_g\} | \{z_g\}, H_0) = \prod_{j=1}^{N_{\rm gal}} {\cal L}_{\rm red}(\hat{z}^j_g | z^j_g),
\end{equation}
and the prior on the galaxy redshifts (ignoring clustering) is
\begin{equation}
p(\{z_g\}) = \prod_{j=1}^{N_{\rm gal}} p_{\rm bg} (z^j_g)
\end{equation}
which we assume to be independent of the prior on $H_0$, so the joint prior on the population level parameters is $p(\{z_g\},H_0) = p(\{z_g\}) p(H_0)$. The joint likelihood for the EM observations comprising the catalogue and the GW data is
\begin{eqnarray}
p(x_i, \{\hat{z}_g\}|H_0,\{z_g\}) &=& {\cal L}_{\rm GW}(x_i|\{z_g\},H_0) 
 \times \nonumber \\ &&{\cal L}_{\rm EM}(\{\hat{z}_g\} | \{z_g\}, H_0)     
\end{eqnarray}
from which we obtain the posterior on the population parameters, $(H_0, \{z_g\})$, via Bayes' theorem
\begin{eqnarray}
p(H_0, \{z_g\} | x_i, \{\hat{z}_g\}) &\propto& \frac{1}{N_{\rm gal}} \left[\sum_{j=1}^{N_{\rm gal}} {\cal L}_{\rm GW}(x_i | d_L(z_g^j, H_0))\right] \times \nonumber \\ && \prod_{j=1}^{N_{\rm gal}} \left[ {\cal L}_{\rm red}(\hat{z}_g^j | z_g^j) p_{\rm bg} (z_g^j) \right].
\end{eqnarray}
We can now integrate out the unknown true galaxy redshifts, $\{ z_g\}$. For each term in the sum over galaxies entering the GW likelihood there is precisely one of these integrals that is over the corresponding galaxy redshift. The other integrals are independent of the GW data and reduce to the evidence for the EM observation of that galaxy
\begin{equation}
{\cal Z}_j = \int {\cal L}_{\rm red}(\hat{z}_g^j | z_g^j) p_{\rm bg} (z_g^j) {\rm d}z_g^j.
\end{equation}
We deduce that the marginalised posterior takes the form
\begin{widetext}
\begin{align}
p(H_0 | x_i, \{\hat{z}\}) &\propto \frac{1}{N_{\rm gal}} \left(\sum_{j=1}^{N_{\rm gal}} \left[ \prod_{k \neq j} {\cal Z}_k\right]  \int {\cal L}_{\rm GW}(x_i | d_L(z_g^j, H_0)) {\cal L}_{\rm red}(\hat{z}_g^j | z_g^j) p_{\rm bg} (z_g^j) \, {\rm d}z_g^j \right) \nonumber \\
&=\frac{1}{N_{\rm gal}} \left[ \prod_{k=1}^{N_{\rm gal}} {\cal Z}_k\right] \sum_{j=1}^{N_{\rm gal}} \left(\int {\cal L}_{\rm GW}(x_i | d_L(z_g^j, H_0)) \frac{{\cal L}_{\rm red}(\hat{z}_g^j | z_g^j) p_{\rm bg} (z_g^j)}{{\cal Z}_j} \, {\rm d}z_g^j \right) \nonumber \\
&=\frac{1}{N_{\rm gal}} \left[ \prod_{k=1}^{N_{\rm gal}} {\cal Z}_k\right] \sum_{j=1}^{N_{\rm gal}} \left(\int {\cal L}_{\rm GW}(x_i | d_L(z_g^j, H_0)) p_{\rm red}(z_g^j | \hat{z}_g^j) \, {\rm d}z_g^j \right) \nonumber \\
&\propto \int {\cal L}_{\rm GW}(x_i | d_L(z, H_0)) \left[ \frac{1}{N_{\rm gal}} \sum_{j=1}^{N_{\rm gal}} p_{\rm red}(z_j | \hat{z}_j) \right] \, {\rm d}z =\int {\cal L}_{\rm GW}(x_i | d_L(z, H_0)) p_{\rm CBC}(z) \, {\rm d}z 
\end{align}
\end{widetext}
and so we recover the result derived earlier. However, this alternative way of deriving the posterior reveals two important corrections that are in principle present, but negligible in practice.

Firstly, we have ignored GW selection effects in the above. These are straightforward to include by replacing ${\cal L}_{\rm GW}$ by ${\cal L}_{\rm GW}/p_{\rm det}(H_0,\{z_g\})$. The detection probability, $p_{\rm det}(H_0,\{z_g\})$, is the integral of the GW likelihood over data sets deemed above the threshold and hence included in the analysis, i.e.,
\begin{eqnarray}
    p_{\rm det}(H_0,\{z_g\})&=&\frac{1}{N_{\rm gal}} \sum_{j=1}^{N_{\rm gal}} \int_{-\infty}^{d_{L}^{\rm thr}} {\cal L}_{\rm GW}(x| d_L(z^j_g, H_0)) \de x \nonumber \\ &=& \frac{1}{N_{\rm gal}} \sum_{j=1}^{N_{\rm gal}}  p_{\rm det}^{\rm GW}(d_L(z^j_g, H_0)). \label{eq:pdet_truez}
\end{eqnarray}
However, this is a function of the {\it true} values of the galaxy redshifts. This dependence of the denominator of the GW likelihood on $\{z_g\}$ breaks the separability of the integrals that we exploited above, unless the true redshifts of the galaxies are perfectly known. In this case, it can be seen that the full hierarchical likelihood reduces to Eq.~\ref{Eq:liknouncert}. In practice, however, we will not know the true redshifts of the galaxies. The detection probability is effectively an average of the galaxy redshift distribution over the whole volume within which GW sources can be observed. If that volume is sufficiently large, as is the case for current GW detectors, then the average galaxy redshift will be approximately uniform in comoving volume, and so the dependence of the detection probability on the actual galaxy redshifts will be relatively weak and so this term can be factored out, reducing the result to the simpler expression used in this paper.

The second correction arises when considering multiple GW observations. The likelihood is the product of likelihoods for each GW observation, but each one of those likelihoods includes the sum over galaxies. This will introduce cross terms which are not simply the product of the individually-marginalized likelihoods, but marginals of the product of the likelihoods over the true redshift of the galaxies. These terms represent corrections for the case when multiple GW events are observed from the same galaxy. In practice, these corrections are small, as there are typically $1/N_{\rm gal}$ times fewer of them than the dominant independent-host terms. These terms will only become important once a significant number of sources in the catalogue share a host. With typical merger rates of a few tens per Gpc$^3$ per year and approximately one galaxy per cubic megaparsec, the typical spacing of mergers in any given galaxy is millions of years, so we will essentially never be in a regime where these corrections matter.\footnote{Space is big. Really big. You just won't believe how vastly hugely mind-bogglingly big it is.} Once again, we emphasise that these terms only arise when the galaxy redshift measurements are imperfect. When galaxy redshifts are known perfectly, Eq.~\ref{Eq:liknouncert} is still exact when analysing many GW sources.

To conclude this section, we note that the inconsistency described here does not represent a fundamental limitation of the galaxy catalogue approach. The analysis can be done consistently by computing the posterior using the full expression
\begin{align}
&\hspace{-0.76cm}p(H_0 | \{x\}) \propto p(H_0) \int d\{z_g\} \left[ \prod_{j=1}^{N_{\rm obs}} \frac{{\cal L}_{\rm GW}(x_j | \{z_g\}, H_0)}{p_{\rm det}(H_0, \{z_g\})} \right] \nonumber \\
&\hspace{3cm} \times \left[ \prod_{k=1}^{N_{\rm gal}} {\cal L}_{\rm red}(\hat{z}^k_g | z_g^k) \right] p_{\rm bg}(\{z_g\}) \nonumber \\
&\hspace{-0.75cm}{\cal L}_{\rm GW}(x_j | \{z_g\}, H_0)=\frac{1}{N_{\rm gal}} \sum_{i=1}^{N_{\rm gal}} {\cal L}_{\rm GW}(x_j | d_L(z^i_g, H_0))\nonumber \\
\label{eq:full_like}
\end{align}
and $p_{\rm det}(H_0, \{z_g\})$ is given in Eq.~(\ref{eq:pdet_truez}). This expression involves multiplication of a sum over galaxies for each GW event, followed by an integration over all the unknown true galaxy redshifts. This is extremely computationally expensive, which is why it is better to use the standard approximate expression, especially since the latter is expected to be accurate in any application to real data. Nonetheless, to further illustrate this issue and its resolution, we will show a simplified example of the application of the full-likelihood framework in Sec.~\ref{sec:onegal}.

\section{Results\label{sec:results}}

In this section we present results on $H_0$ using the statistical framework discussed in Sec.~\ref{sec:method}. We start by describing the simulation of mock data in Sec.~\ref{sec:MDC}, and in Sec.~\ref{sec:200sig} we show results on $H_0$ using the statistical framework in the limit that the number density of galaxies is higher than the number density of GWs. While in Sec.~\ref{sec:onegal} we simulated a case where the number density of galaxies is lower than the number density of mergers.

\subsection{Simulating the Mock data}
\label{sec:MDC}

We now describe step-by-step how we create the mock data. For this mock study, we use galaxies taken from the MICEcat \texttt{v1.0},\footnote{\url{http://maia.ice.cat/mice/}} the Grand Challenge \citep{Hoffmann:2014ida,Carretero:2014ltj,Fosalba_2014,Crocce_2015,Fosalba_2015}. The MICEcat simulation is a lightcone simulation covering 1/8th of the sky out to a redshift of 1.4, down to halo masses of  $2.2 \times 10^{11} h^{-1}~M_\odot$ with a total of about $205$ million galaxies. The fiducial cosmological model in MICEcat \texttt{v1.0} is a Flat $\Lambda$CDM model with cosmological parameters: $H_0= 70$ km/s/Mpc, $\Omega_m=0.25$, $\Omega_b=0.044$, $\Omega_\Lambda= 0.75$, $n_s= 0.95$, $\sigma_8=0.8 $. We also choose this model for our simulations. We note that MICEcat was previously used in \cite{fishbach}, which showed explicitly that dark siren estimations of $H_0$ are unbiased. Below we highlight how MICEcat is employed for this work. 

When generating our mock data, we follow the method in TH21 as closely as possible. We take two sky directions, referred to as ``Direction 1'' and ``Direction 2'', and for each of these, we select galaxies found in an opening angle from the line-of-sight of $\theta=$ 1 and 5 deg, leading to four different scenarios. To simplify the notation we adopt the nomenclature $D_{15}$ to indicate Direction 1 with an opening angle of 5 deg, and similarly for the other combinations of directions and opening angles. On top of this choice, to reduce the overall number of galaxies being considered, we also subsample the galaxies in Direction 2 by half. 

To simulate GW events, given $N_{\rm gal}$ galaxies with a set of \textit{true} redshifts $\{z_g\}$, we randomly choose $N_{\rm ev}$ of them in which to simulate a GW signal. We decide to draw GWs only from galaxies with true redshift below $z_{\rm draw}<1.4$. This is reasonable because the luminosity distance at this redshift ($\sim 10$~Gpc) is much larger than our threshold distance for the GW events. The probability of detecting a GW event from that distance, even when allowing for fluctuations due to the difference between measured and true luminosity distance, is negligible. For each GW signal, we compute the true luminosity distance $d^i_L$ by converting the true redshift using the fiducial MICEcat cosmological parameters. Using the likelihood in Eq.~\ref{eq:GWlikeli}  we draw an observed value for the luminosity distance $\hat{d}^i_L$ following a Gaussian centered at $d_L^i$ with $\sigma=Ad^i_L$. If $\hat{d}^i_L$ is lower than a threshold of $\hat{d}^{\rm thr}_L = 1550$ Mpc (chosen to approximately match the detectability threshold used in TH21), we label the signal as detected and can be used for cosmological inference. 

A crucial aspect of the simulation is the choice of $z_{\rm draw}$ in comparison to the choice of  $\hat{d}^{\rm thr}_L$. Following TH21, we might be tempted to simulate GW events only from galaxies with redshift below a $z_{\rm draw}$ value obtained converting $\hat{d}^{\rm thr}_L$ to a redshift for given cosmological parameters ($H_0$ and $\Omega_m$). This procedure would allow us to detect only GW sources with a \textit{true} luminosity distance below the detection range, and this is inconsistent with the assumed framework to correct for selection biases.
As we argued in Fig.~\ref{fig:GW_detection_prob}, we would expect to find a significant number of sources that are detected even if their \textit{true} luminosity distance is higher than $\hat{d}_L^{\rm thr}$ (the detection range). 
On the contrary, by selecting $z_{\rm max}=1.4$, we are able to simulate GWs at high redshift that become detectable due to fortuitous noise fluctuations. Note that we can always make the choice of injecting GW events in $0<z<0.3$, but we will need to account for this choice, using a $p_{\rm rate}(z)=0$ if $z>0.3$, that must be consistently implemented in the statistical framework. In our analysis, we include the fact that $p_{\rm rate}(z)=0$ for $z>1.4$, in order to avoid introducing any possible bias.

\begin{figure*}[htp!]
    \centering
    \includegraphics{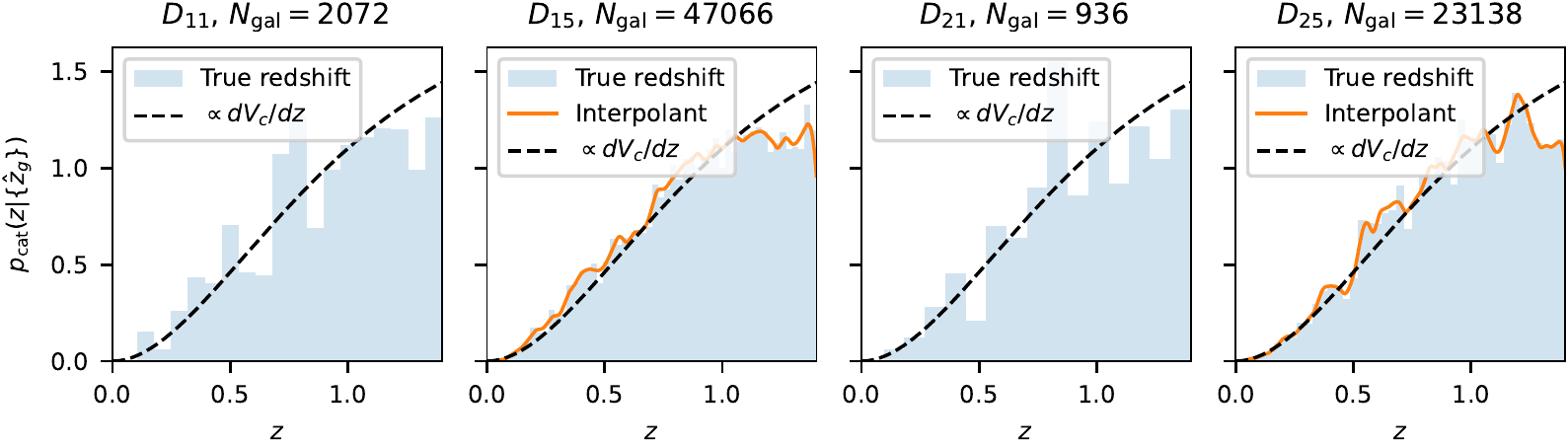}
    \caption{Reconstructions of the galaxy density profile $p(z|C)$ for each of the populations considered in the mock data. The blue histograms indicate the true redshift of the galaxies,  the orange line is the reconstruction of the density profile using Eq.~\ref{eq:pofzc}, and the black dashed line marks a uniform in comoving volume distribution. For the LOSs with small galaxy numbers, we do not fit the redshift interpolant as these are used only in the limit that we perfectly know the redshift (in order to not introduce a bias due to the breaking of the statistical framework).}
    \label{fig:pz}
\end{figure*}

The final step of the simulation is to calculate $p_{\rm CBC}(z)$ to use for the inference in the cases for which we assume that the galaxy redshifts are not perfectly known.
To do so, we draw a set of observed redshifts $\{\hat{z}_g\}$ from the true ones $\{z_g\}$, according to the likelihood model in Eq.~\ref{eq:redlikeli}. By using Eq.~\ref{eq:redpos}, we build a posterior on redshift for every galaxy, which is later used to build the approximant for $p_{\rm CBC}(z)$. To build the approximant, we use a uniform in comoving volume prior, note that this prior does not depend on $H_0$, namely 
\begin{equation}
p_{\rm bg}(z) = \frac{\frac{dV_c}{dz}}{\int_0^\infty \frac{dV_c}{dz} dz},
\end{equation}
where the $H_0$ dependence cancels out, and it depends only on the value of $\Omega_m=0.25$ considered in the  MICEcatv1 simulation.

In Fig.~\ref{fig:pz} we show the reconstruction of the galaxy density profile as a function of redshift for the different lines-of-sight (LOS). We can see that the constructed interpolant for $p_{\rm CBC}$ tracks the true galaxy density profile of the catalog.

\subsection{$H_0$ Inference in the low-galaxy density limit}
\label{sec:200sig}

Following TH21 we simulate 200 GW dark sirens for each of the four directions $D_{11}, D_{15}, D_{21}$ and $D_{25}$ with 3 different scenarios for the error on the GW likelihood: $\sigma_{d_L}/d_L=10\%, 20\%$ and $30\%$.  For this case, we will assume that we perfectly know the redshift of the galaxies. We will use the hierarchical likelihood in Eq.~\ref{Eq:liknouncert}, which we have shown to be formally correct even in the low-galaxy density limit which happens for $D_{11}$ and $D_{21}$.
\begin{figure}
    \centering
    \includegraphics{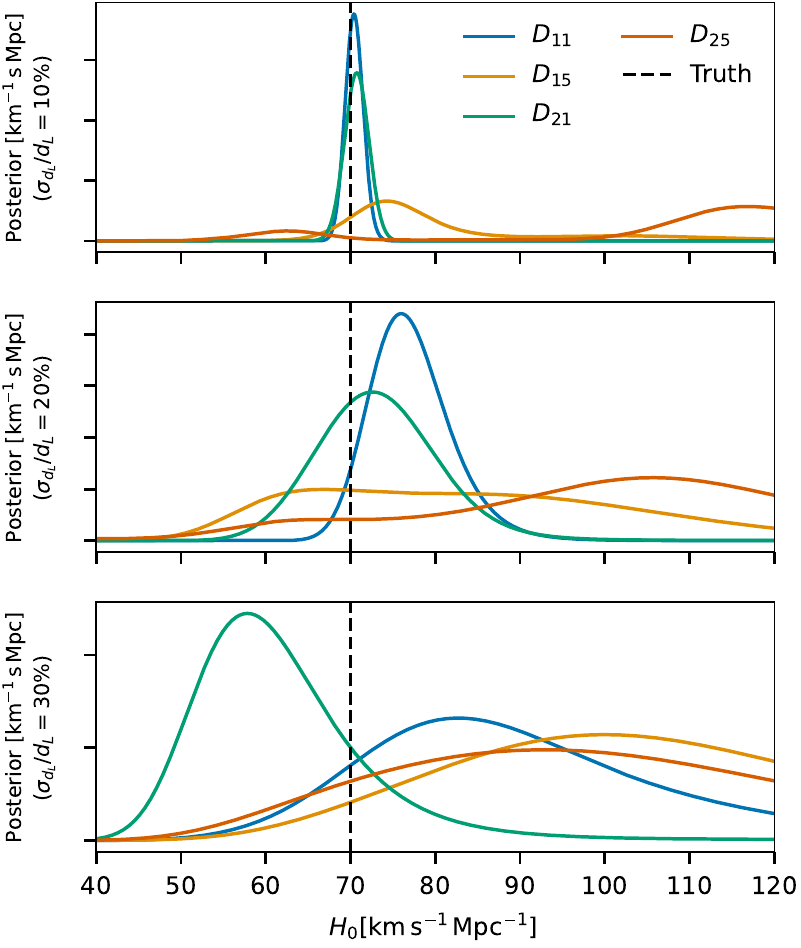}
    \caption{Posteriors of $H_0$ derived for four different lines of sight ($D_1$ and $D_2$) in the MICECat simulations, and for two different sky areas (1$^\circ$ and 5$^\circ$ opening angle for $D_{1x}$ and $D_{2x}$, respectively) around those directions. The top, middle, and bottom panels assume a standard deviation on the luminosity distance which is 10, 20, and 30\% of the luminosity distance, respectively. All the posteriors are the result of combining 200 GW events and assuming galaxy redshift is perfectly known. We do not find a bias. Note that any particular choice of the random seed used for the simulation will lead to fluctuations of the posteriors around the true value. However, by performing several simulations one should find that these fluctuations are compatible with the confidence levels at which the true value is found (see the discussion about PP plots in the text).}
    \label{fig:big_fig_acc_redshift}
\end{figure}

\begin{figure}
    \centering
    \includegraphics{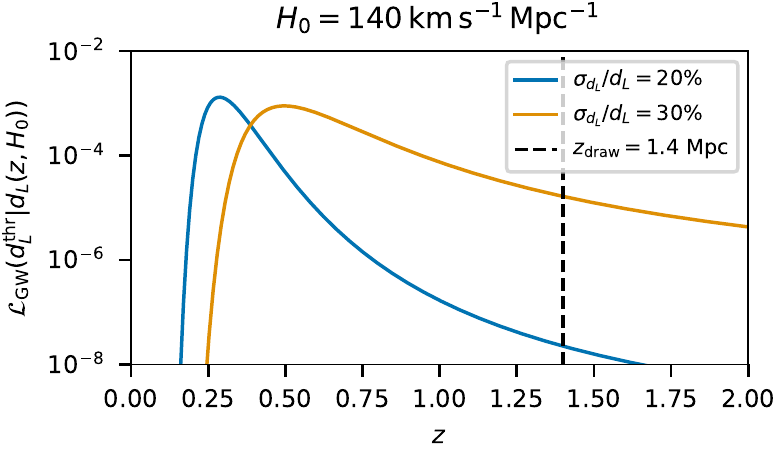}
    \caption{GW likelihood (vertical axis) as a function of the redshift (horizontal axis) for $H_0=140$ km/s/Mpc. The lines show the case of $\sigma_{d_L}/d_L=20\%$ (blue line) and $\sigma_{d_L}/d_L=30\%$ (orange line). The vertical dashed line indicates $z_{\rm draw}$.}
    \label{fig:GW_likelihood}
\end{figure}

In Fig.~\ref{fig:big_fig_acc_redshift} we show the $H_0$ posteriors that we obtain for each line-of-sight, and for each error budget of the GW likelihood. From the plot, we can see that the lines of sight with more galaxies correspond to less constraining power on $H_0$. This is expected, as the GW localization volume will include a larger number of galaxies, and the effect of structures in the galaxy distribution (that can help to provide useful redshift information) is reduced. 
Interestingly, we note that for the cases with $\sigma_{d_L}/d_L =30\%$ it is possible to obtain a posterior which is slightly more informative than the case of $\sigma_{d_L}/d_L=20\%$. This might be counter-intuitive but it is consistent with the fact that for the case of $\sigma_{d_L}/d_L =30\%$ some GW events might be generated at the redshift edge of our simulation $z_{\rm draw}=1.4$. This effectively introduces another redshift scale (an edge) to the inference, which then can help in inferring $H_0$. This effect can be visualized from the GW likelihood. In Fig.~\ref{fig:GW_likelihood}, we plot the GW likelihood as function of redshift for $H_0=140$ km/s/Mpc for a signal detected at $d_{\rm thr}$ with $\sigma_{d_L}/d_L=20\%$ and $\sigma_{d_L}/d_L=30\%$. We can see that the GW likelihood for $\sigma_{d_L}/d_L=30\%$ has some non-negligible support beyond $z_{\rm draw}=1.4$, which is excluded from our statistical analysis. This adds extra information and allows us quickly to exclude high values of $H_0$ as shown in Fig.~\ref{fig:big_fig_acc_redshift}.

\begin{figure}
    \centering
    \includegraphics{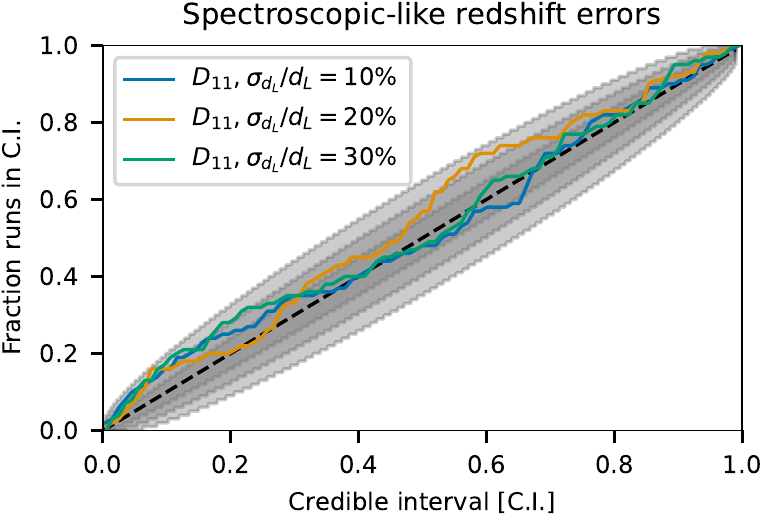}
    \caption{PP-plot generated for the Direction 1 considered here and a nominal error on $d_L$ of 20\% to 30\%. The shaded area shows the 1, 2, 3 $\sigma$ fluctuations of the pp plot around the ideal case. Our simulations show that we are able to recover the input value of $H_0$ without incurring any significant bias.}
    \label{fig:ppplot_spectra}
\end{figure}

To show more quantitatively that the statistical method for dark sirens does not present any significant bias, we generate a probability-probability (PP) plot. PP plots are used to test that parameters subject to Bayesian inference (in our case $H_0$) are consistently recovered. To generate a PP plot, we repeat the $H_0$ inference for 200 GW signals 100 times, each time drawing an injected $H_0$ value between the explored prior range of $[20,140]$ km/s/Mpc. We then check in what credible intervals the true value for each simulation is found. If the analysis is performed properly, we expect to see that e.g. 40\% of injections are found in the 40\% credible intervals. Fig.~\ref{fig:ppplot_spectra} shows the PP plot for our mock data with two lines-of-sight and varying the error budget on $d_L$. As we can see from the plot, the Bayesian framework is suitable to perform the $H_0$ inference as in all cases we are diagonal within $3\sigma$.

\begin{figure}
    \centering
    \includegraphics{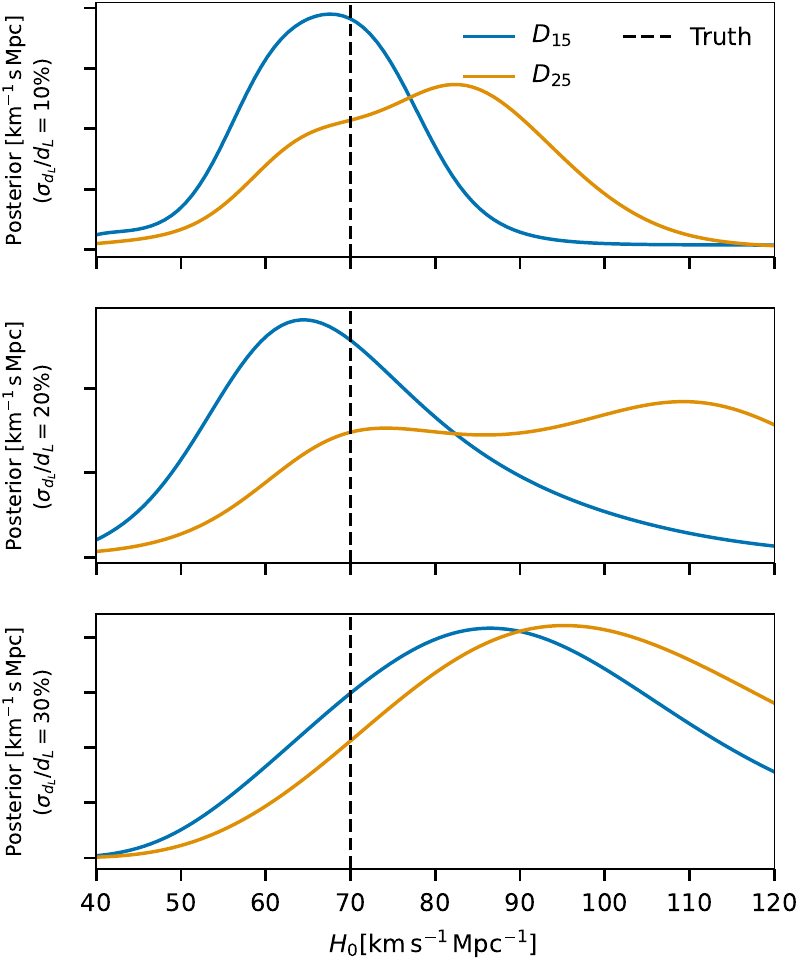}
    \caption{Posteriors of $H_0$ derived for four different lines of sight ($D_1$ and $D_2$) in the MICECat simulations, and for two different sky areas (1$^\circ$ and 5$^\circ$ opening angle for $D_{1x}$ and $D_{2x}$, respectively) around those directions. The top, middle, and bottom panels assume a standard deviation on the luminosity distance which is 10, 20, and 30\% of the luminosity distance, respectively. All the posteriors are the result of combining 200 GW events and assume errors on galaxy redshift. We do not notice any significant bias.}
    \label{fig:big_fig}
\end{figure}

\begin{figure}
    \centering
    \includegraphics{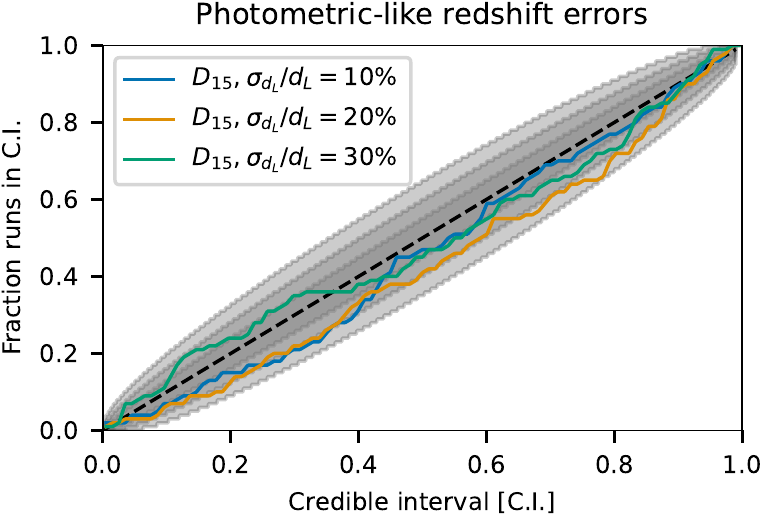}
    \caption{PP-plot generated for Direction 1 and a nominal error on $d_L$ of 20\% to 30\%. The shaded area shows the 1, 2, 3 $\sigma$ fluctuations of the PP plot around the ideal case. Our simulations show that we are able to recover the input value of $H_0$ without incurring any significant bias.}
    \label{fig:ppplot_photo}
\end{figure}

Moreover, we have performed simulations also assuming errors on the galaxy redshifts. In the limit that we have $N_{\rm GW} \geq N_{\rm gal}$, the simplified framework  cannot be used\footnote{This an extremely unrealistic limit where we assume that multiple GW sources are hosted in the same galaxy.} e.g. cases $D_{11}$ and $D_{21}$, otherwise, there could be a bias (see Sec.~\ref{sec:onegal}) for more details. Therefore, we restrict the simulation to $D_{15}$ and $D_{25}$. Figs.~\ref{fig:big_fig} and \ref{fig:ppplot_photo} show a  sample of $H_0$ posteriors and the PP-plots generated from these cases. We can observe that the statistical dark sirens method is unbiased for this case.

\subsection{The one galaxy limit}
\label{sec:onegal}

\begin{figure}
    \centering   
     \includegraphics{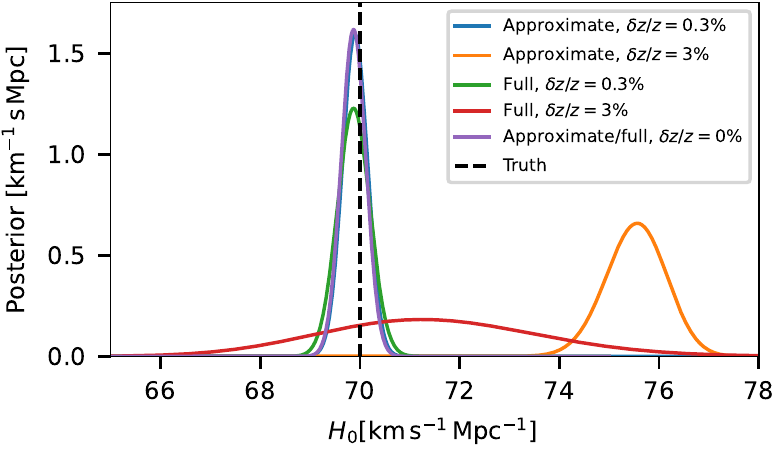}
    \caption{Comparison between posteriors obtained in the one galaxy limit, with $200$ observed gravitational wave sources, computed using the standard approximate likelihood given in Eq.~(\ref{eq:hierarchica}) or using the full likelihood given in Eq.~(\ref{eq:onegal_full_like}). We show results for fractional errors in the $EM$ measurement of the galaxy redshift of $\delta z/z = 0.3\%$ and $3\%$, and also the perfect measurement limit $\delta z/z = 0\%$.}
    \label{fig:postcomp_onegal}
\end{figure}

\begin{figure}
    \centering    
\includegraphics{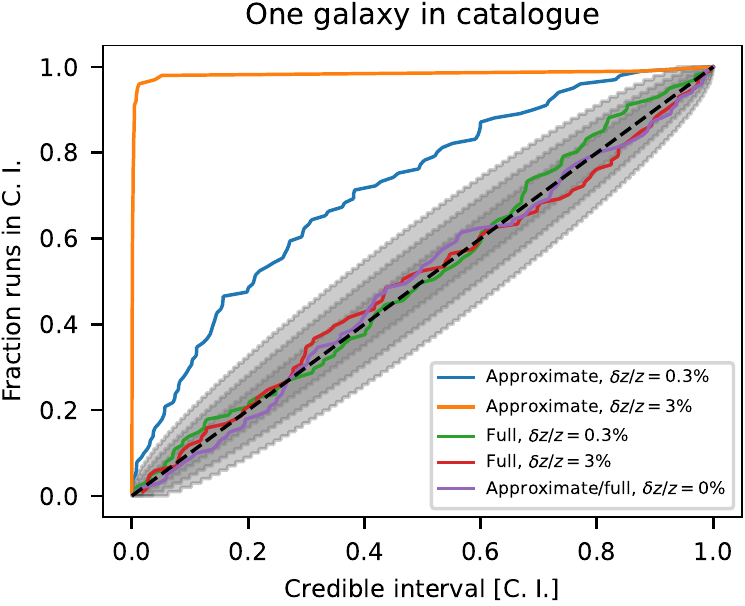}
    \caption{PP-plots for the one galaxy analysis, with $200$ observed gravitational wave sources, comparing the results using the approximate and full likelihoods. We consider the same values of the fractional error in the redshift measurement that were shown in Figure~\ref{fig:postcomp_onegal}.}
    \label{fig:ppplot_onegal}
\end{figure}

As described in Section~\ref{sec:more}, the standard analysis makes some approximations that are only valid in the limit that the gravitational wave detection volume contains a sufficiently large number of galaxies. To illustrate this, we consider here an extreme and unrealistic example in which the galaxy catalogue contains only one galaxy, the redshift of which is known imperfectly from electromagnetic observations. As in previous cases, we generate a catalogue of $200$ GW sources and construct the posterior on $H_0$ using the standard analysis that employs the approximate likelihood. In Figure~\ref{fig:postcomp_onegal} we show example posteriors for catalogues generated with $H_0=70$km s$^{-1}$ Mpc$^{-1}$, assuming the uncertainty in the EM redshift measurement is $\delta z/z = 3\%$ or $0.3\%$. We see that with a $3\%$ error, the posterior is shifted significantly to the right and shows a large bias. For the smaller redshift uncertainty the posterior appears to be unbiased, but in fact a small bias is still present on average. This is revealed by constructing a PP-plot over $100$ random realisations of the Universe. The PP-plot is shown in Figure~\ref{fig:ppplot_onegal} and there is a clear and significant departure from the diagonal, even when $\delta z/z = 0.3\%$.

These biases arise from the fact that in the standard analysis the approximation has been used that the true redshift of the galaxy can be marginalised out of the likelihood separately for each GW event. To carry out a consistent analysis we must use the full likelihood given in Eq.~(\ref{eq:full_like}). In the case that there is only one galaxy, with unknown true redshift $z$ and observed redshift $\hat{z}$, this simplifies to
\begin{align}
&\hspace{-0.76cm}p(H_0 | \{x\}) \propto p(H_0) \int {\rm d} z \left[ \prod_{j=1}^{N_{\rm obs}} \frac{{\cal L}_{\rm GW}(x_j | d_L(z, H_0)}{p_{\rm det}(H_0, z)} \right] \nonumber \\
&\hspace{3cm} \times \left[ {\cal L}_{\rm red}(\hat{z} | z) \right] p(z) 
\label{eq:onegal_full_like}
\end{align}
The results from analysing the same datasets with the full likelihood are also shown in Figures~\ref{fig:postcomp_onegal} and~\ref{fig:ppplot_onegal}. We see that the full likelihood broadens the posterior, making it consistent with the true value even in the case of large redshift uncertainties, and it generates a pp-plot that is perfectly consistent with the expected diagonal line. The figures also show results computed in the perfect-redshift measurement limit, $\delta z/z = 0\%$. In this limit the full and approximate likelihoods are exactly equivalent and so we recover unbiased results using either likelihood.

The reason for going through this example is that it illustrates that it is important to be careful when generating mock catalogues that they have a realistic galaxy density. Using galaxy catalogues that are too sparse can lead to biases because of the inconsistency in the standard analysis that becomes apparent in this limit. However, these are not indicative of a fundamental problem in the analysis, and can be avoided by using the full likelihood of Eq.~(\ref{eq:full_like}).

\section{Discussion} \label{sec:discussion}

We have shown that the dark siren statistical method, utilizing a galaxy catalog of potential hosts, is able to produce unbiased estimates of $H_0$. This is true, however, only if the statistical framework is consistent with the generative model of the population.  An analysis model that is inconsistent with the simulations will incur biases. 

\citet{2021arXiv211200241T} concludes that the bias on $H_0$ is introduced by the absence of galaxy clustering, which is exacerbated when the number of galaxies in a given direction is large (e.g. the cases $D_{15}$ and $D_{25}$), or when the error budget on the GW luminosity distance is large (leading to the clustering effects being washed out in the GW localization volume). We argue that the conclusions in \cite{2021arXiv211200241T} are likely due to an inconsistent procedure for the generation of their mock data and calculating the selection effects for their GW events. The following sections will demonstrate that an absence of structure in the redshift distribution of galaxies does not lead to a biased estimate of \ac{H0} (indeed the result merely becomes uninformative), and that similar biases to those seen in TH21 can be produced by introducing inconsistencies between the mock data generation and analysis.

\subsection{The absence of clustering}

In Sec.~\ref{sec:results} we have shown that in the limit that there are many galaxies along the line-of-sight (e.g $D_{15}$ and $D_{25}$) we recover an unbiased $H_0$ result.  As discussed above, in this limit we expect the constraining power on $H_0$ to significantly decrease. It can be demonstrated mathematically, as we do in App.~\ref{app:A}, that under some simplified assumptions the $H_0$ posterior is expected to be flat (assuming a flat prior) in this case. 

To further demonstrate this, here we perform a full simulation of this measurement.
\begin{figure}
    \centering
    \includegraphics{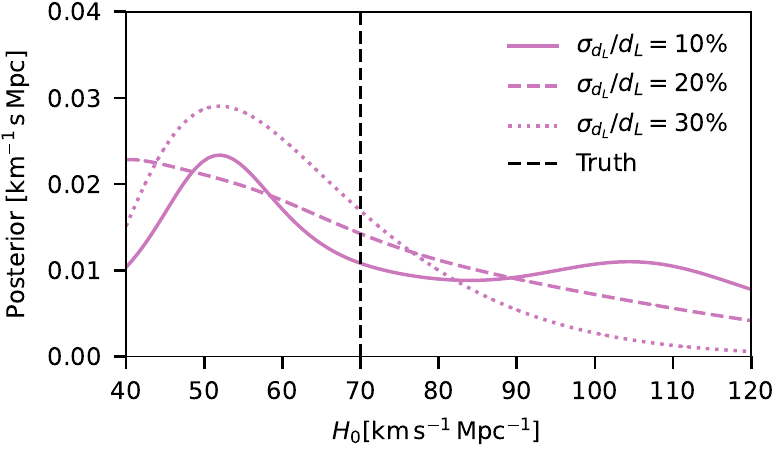}
    \caption{$H_0$ posteriors for 200 GW events drawn from a uniform in comoving volume distribution of $10^6$ galaxies. The several lines indicate various fractional errors on $d_L$ for the GW likelihood model. The absence of clustering significantly weakens the result, rendering it nearly uninformative in the limit of large numbers of galaxies, but does not incur in significant biases.}
    \label{fig:big_fig_realuni}
\end{figure}
We repeat the steps of simulating a Mock Data Challenge (MDC) explained in Sec.~\ref{sec:MDC} with the only difference that the galaxy distribution is synthetically generated to be uniform in comoving volume without any large-scale clustering. We draw $2\times 10^6$ galaxies for this case in order to have a distribution as continuous as possible, and we use the same uncertainty model for the observed redshifts of the galaxies. In other words $p_{\rm gal}(z) \propto \frac{dV_c}{dz}$. In Fig.~\ref{fig:big_fig_realuni}, we show the $H_0$ posteriors of 200 GW signals combined for a relative error on $d_L$ on the GW likelihood model of $10\%, 20\%$ and $30\%$. We note that the recovered $H_0$ posterior is not as constraining as in the previous cases, as expected. The posterior does not display any noticeable bias in this case. Therefore, contrary to \citet{2021arXiv211200241T}, we conclude that the absence of clustering is not expected to introduce a bias on the $H_0$ estimation. In reality, the Universe is \emph{not} uniform in comoving volume on smaller scales, as we know that the Universe's large-scale structure does exhibit clustering, but the effect of clustering is to enhance the \ac{H0} constraint, rather than being essential to it. 

\subsection{Possible sources of inconsistency in implementing the standard siren method }

In this section we speculate of possible inconsistencies between the statistical framework and the generation of mock data can easily introduce biases in the estimation of $H_0$. It is difficult to characterize the interplay of different issues, and how they might translate to a final $H_0$ bias, for this reason, we explore one possible inconsistency at a time.
In what follows we focus on a number of possible errors and demonstrate how they impact the $H_0$ posterior. 

\begin{figure}
    \centering
    \includegraphics{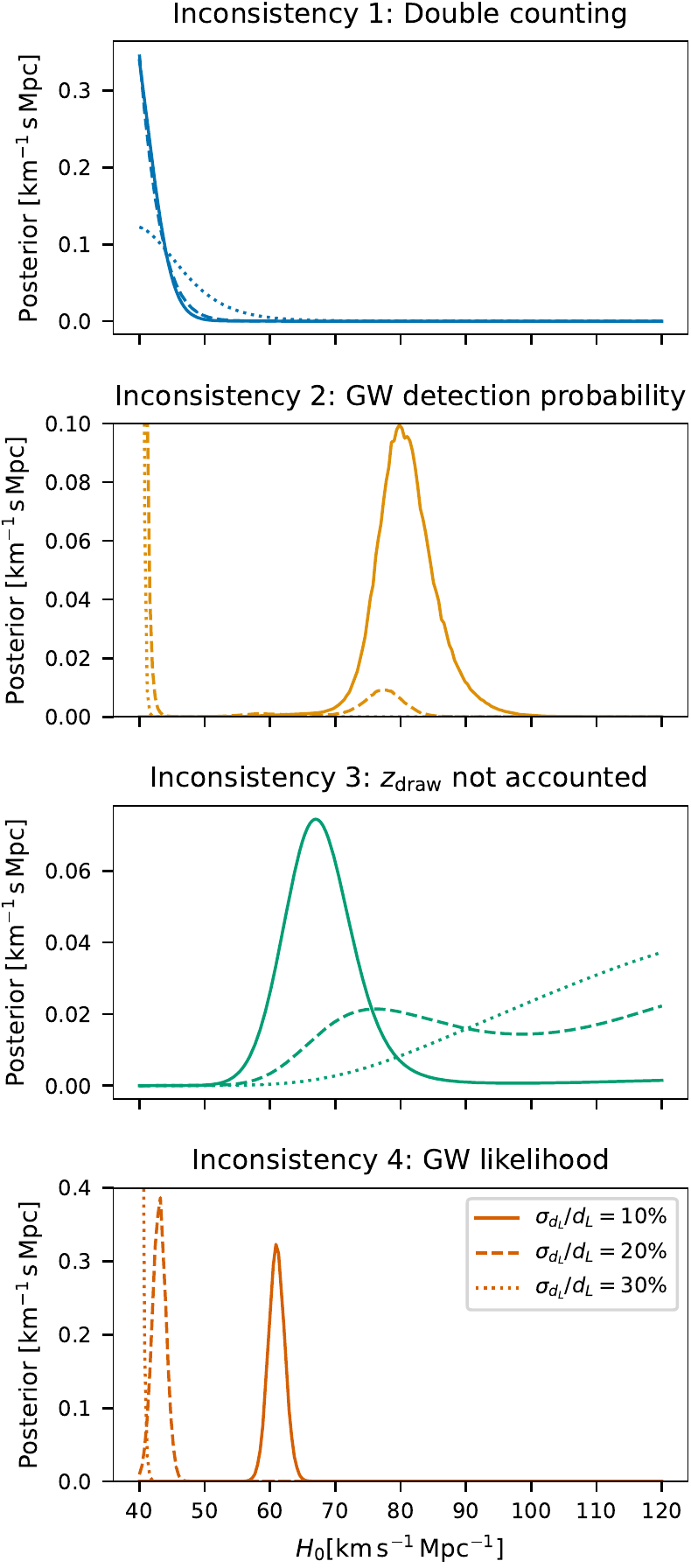}
    \caption{$H_0$ posteriors with 200 GW events for several cases with inconsistent choices in either the simulation of mock data or the statistical framework, for the $D_{15}$ line-of-sight. The solid, dashed and dotted lines are obtained with $\sigma_{d_L}/d_L=10\%, 20\%$ and $30\%$. First panel: GW events are generated by counting twice over the galaxy catalog. Second panel: The detection probability in the analysis is set as a Heaviside step function. Third panel: GW events are drawn from $z<0.3$ and this correction is ignored in the analysis. Fourth panel: The luminosity distance dependence of the GW likelihood standard deviation is not properly taken into account in the formalism. Note that for this particular case alone we show $D_{21}$, which shows a more dramatic behaviour than $D_{15}$, at least for the specific random seeds we tested.}
    \label{fig:inconsistencies}
\end{figure}

\paragraph{Inconsistency 1, Double counting} A first issue is assigning GW events to the sample of MICEcat galaxies using a weight factor similar to Eq.~\ref{eq:pofzc}. However, since the sample of MICEcat galaxies is already distributed as Eq.~\ref{eq:pofzc}, this procedure effectively introduces a ``double counting'' problem. In other words, if the distribution of galaxies follows a $z^2$ profile, the distribution of GW events will follow a $z^4$ profile. This has a crucial impact on the $H_0$ inference when combined with errors in the luminosity distance generation and the presence of a detection threshold.  Understanding the implication of the $H_0$ bias is not trivial for the dark siren approach, therefore we perform a simulation.  In Fig.~\ref{fig:inconsistencies} (top panel), we repeat the analysis as in Sec.~\ref{sec:results} but generating GWs in galaxies with an extra weight factor given by Eq.~\ref{eq:pofzc} (the analysis is then performed normally). As we see, this type of mismatch usually results in a bias towards lower values of $H_0$. 

\paragraph{Inconsistency 2, GW detection probability}
A second issue is a general misinterpretation of observed quantities (``data'') and latent variables (``true values'') and their interaction with the selection biases. As we detail in Sec.~\ref{sec:method}, the selection bias on the GW side should be corrected using the GW detection probability. The latter is defined by integrating the likelihood over all possible data sets that can be detected given a true value of the luminosity distance. TH21 assumes that the detection probability is a Heaviside step function of the true luminosity distance, with the probability dropping to 0 at $\hat{d}^{\rm thr}_L$. This is correct only in the limit that  there are no errors on the luminosity distance estimation. Fig.~\ref{fig:inconsistencies} (middle panel) shows that, when the luminosity distance estimation comes with a non-negligible uncertainty, assuming a Heaviside step function for the detection probability can bias the estimation of the Hubble constant to lower values.

\paragraph{Inconsistency 3, $z_{\rm draw}$ not accounted}
A third potential misconception is the fact that GW events are drawn from galaxies with $z<0.3$, but this aspect is not accounted for in the analysis. In order to understand the biases introduced by this type of error, we simulate GW events from galaxies at $z<0.3$ and then build the interpolant $p_{\rm CBC}(z)$ for the analysis using $z_{\rm draw}=1.4$. In other words, the analysis accounts also for selection effects due to GW hosts present between $0.3<z<1.4$, while the former simulation only considered GW hosts $z<0.3$. In Fig.~\ref{fig:inconsistencies} (third panel), we show that this type of error might introduce a high $H_0$ bias. We note that in a real analysis, such as that done in \cite{gwtc3_H0}, we do not know if there is a maximum redshift at which GW events are generated. However, this is taken into account by fitting simultaneously with flexible merger rate models alongside the value of $H_0$.

\paragraph{Inconsistency 4, GW likelihood mismodeling:} Another possible source of error is treating the GW likelihood as if the standard deviation was not dependent on the true value of $d_L$, although the simulations are made assuming this dependence (see Section \ref{sec:method}). By dropping the overall normalization factor in the GW likelihood, one is in practice ignoring a part of the likelihood that depends on the true luminosity distance. This causes a biased dependence of the $H_0$ posterior on the luminosity distance uncertainty, which we show in Fig.~\ref{fig:inconsistencies} (bottom panel). We find that in this case the inconsistency has the effect of biasing $H_0$ towards lower values for increasing values of $\sigma_{d_L}$. similarly to what TH21 found. In the bottom panel of Fig. \ref{fig:inconsistencies} we show the results using a 1 deg light cone simulation (instead of the 5 deg $D_{15}$ simulation) as it shows a stronger bias change with $\sigma_{d_L}$. We also assume a Heaviside function for the selection effects as in TH21 so that the $\sigma_{d_L}$ dependence on true luminosity distance is also not taken into account in this term. A general discussion about the misinterpretation of latent and observed variables, and the bias caused by  assuming that observables depend on intrinsic properties in the mocks but not in the likelihood modeling in the context of GWs is also provided in \url{https://github.com/farr/MockPosteriors}. %Given this, and the fact that we are able to produce unbiased constraints in this analysis, we conclude that \sout{the main source of bias found in TH21 (the one that depends on the  luminosity distance uncertainty)} \daniel{is not a fundamental aspect of the dark siren method, nor due to irreducible degeneracies between the parameters at play, but it instead due to a mismodeling of the GW likelihood.}
 %\apa{Check with Will that this link should be here?}

\paragraph{Inconsistency 5, simulating dark siren events along the same line of sight.} The statistical method used in TH21 is for the high-galaxy density regime, but this is inconsistent with the generation of data from $D_{11}$ and $D_{21}$, where it is possible for multiple GW events to be drawn from the same low-redshift host galaxy. Some of the $H_0$ biases for $D_{11}$ and $D_{21}$ might originate from this inconsistency, as we have shown in Sec.~\ref{sec:onegal}. However, in our own analyses, this produced a noticeable bias less frequently than the other potential sources of bias we explored. Moreover, we find that a low-$H_0$ bias will arise if there are low-redshift overdensities along the line of sight and the full likelihood analysis is not used. When simulating multiple events along the same line of sight, if there is an overdensity of galaxies in that direction, the $H_0$ posterior is expected to display a peak at an $H_0$ lower than the input value, that corresponds to the redshift of the overdensity and luminosity distance of the peak of the GW distribution (likely $\sim 1500$ Mpc in this work, close to the detection horizon). Especially when using the light cones with fewer galaxies, there are very few galaxies at the redshift of interest $z\lesssim 0.3$, so we are sensitive to the small number statistics of a peak in the redshift distribution that may shift from realization to realization (as one does not always get the same exact redshift peak, and thus corresponding $H_0$ peak, in all realizations).
As there are significantly more galaxies and volume at larger redshifts than at lower redshifts, which results in the effect of the overdensities on the $H_0$ posterior being typically less pronounced in the former regime rather than in the latter, this effect is only more significant at lower $H_0$. Because in reality it is extremely unlikely to have multiple events from one narrow line of sight, we advise against using a small area light cone simulation to make dark siren analyses. Alternatively, the presence of the same galaxies across multiple GW events needs to be taken into account with the full likelihood analysis so as not to incur in a bias.

\section{Conclusions}

We have presented an introduction to the dark siren statistical method. We have explicitly demonstrated that, contrary to TH21, the dark siren statistical method can robustly recover an unbiased estimate of $H_0$. We expect the method to produce unbiased estimates of additional cosmological parameters.
We have also shown that in the limit of the absence of galaxy clustering, the dark siren statistical method continues to provide an unbiased estimate of $H_0$. Notebooks showing how to reproduce this analysis can be found at \url{https://github.com/simone-mastrogiovanni/hitchhiker_guide_dark_sirens}.

We emphasize that this work is not a forecast, but instead is meant to provide a pedagogical introduction to the dark siren approach, highlighting common pitfalls.
%example to show how the dark siren method works and what common errors could happen. 
Because of the specific assumptions that we have made, this is not a realistic simulation of GW events and their detections, and we do not provide estimates for the future sensitivity of dark siren probes (see e.g. \citealt{2018Natur.562..545C,2020PhRvD.101l2001G} for forecasts). This is why we do not specifically quote precision measurements of $H_0$ throughout the manuscript. 

Moreover, we remind the reader that a data analysis framework able to analyze real GW data is, in some aspects, more complicated than what is discussed in this paper. On the GW side, as has already been noted in the literature, selection effects based on GW sensitivity studies injecting signals in real data should also be carefully taken into account for standard sirens with counterparts (e.g.~\citealt{Mortlock_2019,2019RNAAS...3...66F}). Moreover, assumptions about the underlying population of compact object binaries can have a significant impact on estimating the Hubble constant not only for the galaxy catalog approach \citep{gwtc3_H0}, but perhaps more importantly for a GW-only approach to standard siren cosmology \citep{mastrogiovanni_2021,Mukherjee:2021rtw, Ezquiaga_2022, Karathanasis:2022rtr}. On the galaxy catalog side, there are several challenges related to the completeness correction of the galaxy catalog and the presence of a possible correlation between galaxy intrinsic luminosity and merger rates \citep{2020PhRvD.101l2001G,Gray2022}. Also, techniques exploring spatial clustering using cross-correlation  need to properly mitigate the effects from GW bias parameters and its redshift evolution as demonstrated in \citep{Mukherjee:2020hyn,Mukherjee:2022afz}. Clearly, the full Bayesian framework is more complicated for a statistical dark standard siren analysis as opposed to the case where a unique host galaxy is identified, so it is understandable that this method poses more challenges than the one simulated in this paper.  However, it is important to note that \emph{all} standard siren methods are dependent on the same assumptions and potential sources of mismodeling considered here. 

Another caveat of this analysis, since it is built in part around the assumptions of TH21 to show how not to incur the biases they claim to find, is that of drawing a large number of events from the same line of sight. In this analysis, we have used only two directions to simulate all the events of the GW sources. When drawing from the same line of sight, the contribution of the large-scale structure is always the same, so your posterior may prefer the specific values of $H_0$ around the value corresponding to the overdensities along the line of sight for distances close to the peak of the detected GW source distribution, thereby lacking the expected variation in the large scale structure distribution. In reality, over different lines of sight, the contributions from the different under densities/overdensities will cancel out. %However, with the limit of the number of GW detections being lower than the number of galaxies in the simulation, it appears that this does not give rise to noticeable biases. 
We caution the reader from running the simulations we have made available using a number of GW events much larger than the number of galaxies.

To summarize, our analysis shows that using the dark siren statistical method, we can measure the value of the Hubble constant \ac{H0} without any bias, in contrast to the recent claim by TH21. Accurate measurements of the Hubble constant require correctly taking into account the selection effects in both the galaxy catalog side and GW side, and a proper modeling of the likelihoods in question. We encourage research studies focused on the impact of population assumptions and selection effects to advance the entire field of standard siren cosmology.

\begin{acknowledgements}

We thank Dragan Huterer and Emery Trott for clarifying the details of their analysis and for sharing their galaxy sample from the MICEcat simulation so as to enable us to provide a direct comparison to their work. We thank the Kavli Institute for Cosmological Physics at the University of Chicago for hosting ``The quest for Precision Gravitational Wave Cosmology Workshop'' organized by Jose Maria Ezquiaga and DEH, where part of this work has been discussed. The research of A.~Ghosh is supported by the Ghent University BOF project BOF/STA/202009/040 and the Fonds Wetenschappelijk Onderzoek (FWO) iBOF project BOF20/IBF/124. A.~Palmese acknowledges support for this work was provided by NASA through the NASA Hubble Fellowship grant HST-HF2-51488.001-A awarded by the Space Telescope Science Institute, which is operated by Association of Universities for Research in Astronomy, Inc., for NASA, under contract NAS5-26555.
S.~Mastrogiovanni thanks the Albert Einstein Institute, Potsdam, for the hospitality while this work has been developed. The work of S. Mukherjee is a part of the $\langle \texttt{data|theory}\rangle$ \texttt{Universe-Lab} which is supported by the TIFR and the Department of Atomic Energy, Government of India.
The research of R. Gray is supported by the European Research Council, starting grant SHADE 949572.
N. Tamanini acknowledges support form the French space agency CNES in the framework of LISA.

This work has made use of CosmoHub \citep{TALLADA2020100391,2017ehep.confE.488C}. CosmoHub has been developed by the Port d'Informació Científica (PIC), maintained through a collaboration of the Institut de Física d'Altes Energies (IFAE) and the Centro de Investigaciones Energéticas, Medioambientales y Tecnológicas (CIEMAT) and the Institute of Space Sciences (CSIC \& IEEC), and was partially funded by the "Plan Estatal de Investigación Científica y Técnica y de Innovación" program of the Spanish government.

\end{acknowledgements}

\bibliography{bib}

\appendix

\section{Dark sirens in absence of galaxy clustering: a simplified calculation}
\label{app:A}

In this Appendix, we show that under the assumptions that \textit{(i)} the GW likelihood and detection criteria are only function of the source luminosity distance and \textit{(ii)} we are in the local Universe and the distribution of galaxies is uniform, then we expect to obtain an uninformative $H_0$ likelihood. This is a mathematical proof that the statement in TH21 according to which the absence of galaxy clustering one recovers a biased posterior is incorrect.
Let us remind that the hierarchical likelihood is given by 
\begin{equation}
     \mathcal{L}(\{x\}|H_0) \propto \prod_i^{N_{\rm obs}}  \frac{ \int \de z \mathcal{L}_{\rm GW}(\hat{d}^i_L |d_L(z,H_0))~p_{\rm CBC}(z)}{\int \de z~ P^{\rm GW}_{\rm det}(z,H_0) ~p_{\rm CBC}(z)},
     \label{eq:ll2}
\end{equation}
and the posterior on $H_0$ can be calculated as
\begin{equation}
    p(H_0|\{x\}) = \frac{\mathcal{L}(\{x\}|H_0) p(H_0)}{\int \mathcal{L}(\{x\}|H_0) p(H_0) \de H_0}.
    \label{eq:posa}
\end{equation}

If we are in the local Universe, i.e. distance can be approximated as
\begin{equation}
    d_L \approx \frac{cz}{H_0}
\end{equation}
and if the distribution of galaxies is, we can write it as continuous and it is 
\begin{equation}
    p_{\rm CBC}(z) \approx p_{\rm cat}(z) = \frac{3}{z^3_{\rm max}} z^2.
\end{equation}
In the above Equation, $z_{\rm max}$ is a redshift limit up to which galaxies are generated.
Moreover, if the single GW likelihood only depends on the source luminosity distance, and the detection criteria depends on a threshold luminosity distance $\hat{d}^{\rm thr}_L$ such that $d_L(z_{\rm max},H_0) \gg \hat{d}^{\rm thr}_L$ for all the $H_0$ values explored in the analysis, then the selection bias can be written as
\begin{equation}
    \int  P^{\rm GW}_{\rm det}(z,H_0) ~p_{\rm CBC}(z) \de z=\int  P^{\rm GW}_{\rm det}(d_L)  ~p_{\rm CBC}(z (d_L,H_0)) \left|\frac{\partial z}{\partial d_L}\right| \de d_L = k H_0^3.
    \label{eq:de}
\end{equation}
where $k$ is a constant. To solve the integral in the above Equation, we have just performed a change of variable from $z$ to $d_L$.
We can perform a similar procedure for the numerator of Eq.~\ref{eq:ll2} and show that this is given by
\begin{equation}
    \int \mathcal{L}_{\rm GW}(\hat{d}^i_L|d_L(z,H_0))~p_{\rm CBC}(z)  \de z = \int  \mathcal{L}_{\rm GW}(\hat{d}^i_L|d_L(z,H_0))  ~p_{\rm CBC}(z (d_L,H_0)) \left|\frac{\partial z}{\partial d_L}\right| \de d_L = C_i H_0^3,
    \label{eq:nu}
\end{equation}
where again we have performed a change of variable to solve the integral and $C_i$ is a normalization constant that depends on the signal detected from the data realization.

Note that, when calculating Eq.~\ref{eq:de} and Eq.~\ref{eq:nu} the hypothesis that $d_L(z_{\rm max},H_0) \gg \hat{d}^{\rm thr}_L$ makes sure that $C_i$ and $k$ does not depend on $H_0$. If this condition is not satisfied, these constants will depend on $H_0$ and must be correctly included in the analyis.
By using Eq.~\ref{eq:de} and Eq.~\ref{eq:nu} to calculate Eq.~\ref{eq:ll2}, we obtain that the GW likelihood is constant and does not depend on $H_0$, namely
\begin{equation}
    \mathcal{L}(\{x\}|H_0)=\prod_i^{N_{\rm obs}} \frac{C_i}{k}.
\end{equation}
Therefore, the posterior on $H_0$ equals the prior used in Eq.~\ref{eq:posa}.

\end{document}